\PassOptionsToPackage{hyphens}{url}
\newif\ifarxiv
\arxivtrue
\documentclass[camera,letterpaper,nomarginnotes,nonarrowgutter]{jpaper}
\usepackage{textcomp}
\usepackage{tikz}

\usepackage{cite}
\usepackage{amsmath,amssymb,amsfonts}
\usepackage{graphicx}
\usepackage{textcomp}
\usepackage{xcolor}
\usepackage{fancyhdr}
\usepackage{hyperref}
\usepackage{titlesec}
\usepackage{balance}
\usepackage{multirow}
\usepackage{tabularx}
\usepackage{booktabs}
\usepackage{setspace}
\usepackage{xspace}
\usepackage{color}
\usepackage{listings}
\usepackage{xcolor}
\usepackage{algorithm}
\usepackage{algpseudocode}
\usepackage{xspace}
\usepackage{subcaption}
\algrenewcommand\algorithmicrequire{\textbf{Input:}}
\algrenewcommand\algorithmicensure{\textbf{Output:}}
\definecolor{darkspringgreen}{rgb}{0.09, 0.45, 0.27}
\definecolor{denim}{rgb}{0.08, 0.38, 0.74}
\definecolor{darkolivegreen}{rgb}{0.33, 0.42, 0.18}
\definecolor{tangerine}{rgb}{0.95, 0.52, 0.0}
\definecolor{mahogany}{rgb}{0.75, 0.25, 0.0}

\definecolor{uglyyellow}{rgb}{0.99, 0.93, 0.0}
\definecolor{nbs}{rgb}{0.35, 0.31, 0.81}

\definecolor{darkred}{rgb}{0.75, 0.1, 0.1}

\newcommand\revmid[2]{\todo[linecolor=#1,backgroundcolor=#1!15,bordercolor=#1]{\textcolor{black}{\textbf{#2}}}}

\definecolor{ra}{HTML}{4E79A7}
\definecolor{rb}{HTML}{E07B39}
\definecolor{rc}{HTML}{59A14F}
\definecolor{rd}{HTML}{C44E52}
\definecolor{re}{HTML}{8E6BBE}
\definecolor{cq}{HTML}{2F8F8F}

\definecolor{ra}{HTML}{E6194B}       %
\definecolor{rb}{HTML}{4363D8}       %
\definecolor{rc}{HTML}{3CB44B}       %
\definecolor{rd}{HTML}{E76F51}
\definecolor{re}{HTML}{808000}       %
\definecolor{rf}{HTML}{911EB4}       %

\definecolor{cq}{HTML}{42D4F4}       %
\definecolor{cqtwo}{HTML}{F032E6}    %
\definecolor{cqthree}{HTML}{9A6324}  %
\definecolor{cqfour}{HTML}{F58231}   %
\definecolor{cqfive}{HTML}{469990}   %

\newcommand{\labela}[1]{\phantomsection\label{resp:A#1}\revmid{ra}{A#1}}
\newcommand{\labelb}[1]{\phantomsection\label{resp:B#1}\revmid{rb}{B#1}}
\newcommand{\labelc}[1]{\phantomsection\label{resp:C#1}\revmid{rc}{C#1}}
\newcommand{\labeld}[1]{\phantomsection\label{resp:D#1}\revmid{rd}{D#1}}
\newcommand{\labele}[1]{\phantomsection\label{resp:E#1}\revmid{re}{E#1}}
\newcommand{\labelf}[1]{\phantomsection\label{resp:F#1}\revmid{rf}{F#1}}

\newcommand{\labelcq}[1]{\phantomsection\label{resp:CQ#1}\revmid{cq}{CQ#1}}
\newcommand{\labelcqtwo}[1]{\phantomsection\label{resp:CQ#1}\revmid{cqtwo}{CQ#1}}
\newcommand{\labelcqthree}[1]{\phantomsection\label{resp:CQ#1}\revmid{cqthree}{CQ#1}}
\newcommand{\labelcqfour}[1]{\phantomsection\label{resp:CQ#1}\revmid{cqfour}{CQ#1}}
\newcommand{\labelcqfive}[1]{\phantomsection\label{resp:CQ#1}\revmid{cqfive}{CQ#1}}

\makeatletter
\@ifclassloaded{acmart}{}{\@ifclassloaded{jpaper}{}{%
\usepackage{titlesec}
\titlespacing\section{0pt}{5pt plus 1pt minus 1pt}{2pt plus 1pt minus 1pt}
\titlespacing\subsection{0pt}{5pt plus 1pt minus 1pt}{2pt plus 1pt minus 1pt}
\titlespacing\subsubsection{0pt}{5pt plus 1pt minus 1pt}{2pt plus 1pt minus 1pt}
}}
\makeatother

\makeatletter
\g@addto@macro{\normalsize}{%
  \setlength{\abovedisplayskip}{2pt plus 0.5pt minus 1pt}
  \setlength{\belowdisplayskip}{2pt plus 0.5pt minus 1pt}
  \setlength{\abovedisplayshortskip}{0pt}
  \setlength{\belowdisplayshortskip}{0pt}
  \setlength{\intextsep}{2pt plus 1pt minus 1pt}
  \setlength{\textfloatsep}{2pt plus 1pt minus 1pt}
  \setlength{\floatsep}{2pt plus 1pt minus 1pt}
  \setlength{\skip\footins}{2pt plus 1pt minus 1pt}
  \setlength{\abovecaptionskip}{1pt plus 1pt minus 1pt}
  \setlength{\belowcaptionskip}{1pt}
}
\makeatother

\newcommand\toolname{Motivo}
\newcommand\tgm{\textsf{\toolname}\xspace}

\newcommand\tgmindex{\textsf{\toolname{}Index}\xspace}
\newcommand\tgmext{\textsf{\toolname{}Ext}\xspace}

\newcommand\tgmsw{\textsf{\toolname{}-CPU}\xspace}
\newcommand\tgmcpu{\textsf{\toolname{}-CPU}\xspace}
\newcommand\tgmgpu{\textsf{\toolname{}-GPU}\xspace}

\definecolor{seagreen}{rgb}{0.18, 0.55, 0.34}

\definecolor{darkpink}{rgb}{0.88, 0.28, 0.54}
\definecolor{forestgreen}{rgb}{0.0, 0.27, 0.13}
\definecolor{amber}{rgb}{1.0, 0.49, 0.0}

\newcommand\konkanello[1]{\textcolor{black}{#1}}

\newcommand{\sect}[1]{{§#1}\xspace}
\newcommand{\head}[1]{{\noindent\textbf{#1.}\xspace}} %
\newcommand{\fig}[1]{{Fig.~#1}\xspace} %
\newcommand{\tab}[1]{{Table~#1}\xspace} %
\newcommand{\inum}[1]{(\textit{#1})\xspace}

\usepackage{tikz}

\newcommand{\squishlist}{
 \begin{list}{$\circ$}
  { \setlength{\itemsep}{0pt}
     \setlength{\parsep}{0pt}
     \setlength{\topsep}{3pt}
     \setlength{\partopsep}{0pt}
     \setlength{\leftmargin}{1em}
     \setlength{\labelwidth}{1em}
     \setlength{\labelsep}{0.5em} } }

\newcommand{\squishend}{
  \end{list}  }

\newcommand*\circled[1]{\tikz[baseline=(char.base)]{\node[shape=circle,fill,inner sep=0.5pt] (char) {\textcolor{white}{#1}};}}

\definecolor{denim}{rgb}{0.08, 0.38, 0.74}
\definecolor{azure(colorwheel)}{rgb}{0.0, 0.5, 1.0}
\definecolor{greenp}{rgb}{0.0, 0.65, 0.50}
\definecolor{peach}{rgb}{0.97, 0.51, 0.47}
\definecolor{darkmagenta}{rgb}{0.55, 0.0, 0.55}
\definecolor{royalblue(web)}{rgb}{0.25, 0.41, 0.88}
\definecolor{ao(english)}{rgb}{0.0, 0.5, 0.0}
\definecolor{ForestGreen}{RGB}{34,139,34}

\usepackage[colorinlistoftodos,prependcaption,textsize=scriptsize]{todonotes}

\usepackage[bottom]{footmisc}
\usepackage{dblfloatfix}
\usepackage{array}
 \usepackage{booktabs}

\definecolor{ufogreen}{rgb}{0.1, 0.6, 0.4}
\definecolor{ufogreen}{rgb}{0.1, 0.6, 0.4}

\usepackage{listings}
\usepackage{xcolor}
\usepackage{iftex}
\ifPDFTeX\usepackage[T1]{fontenc}\fi %
\usepackage{zi4} %
\usepackage{caption} %
\usepackage{float} %

\lstdefinestyle{custompseudocode}{
  belowcaptionskip=1\baselineskip,
  breaklines=true,
  xleftmargin=\parindent,
  language=Python,
  showstringspaces=false,
  basicstyle=\small\ttfamily,
  keywordstyle=\bfseries\color{green!40!black},
  commentstyle=\itshape\color{purple},
  stringstyle=\color{orange},
  numbers=left,
  numberstyle=\scriptsize\color{black},
  numbersep=8pt,
  morekeywords={function}, %
  keywordstyle=[2]\bfseries, %
  escapeinside={*@}{@*}, %
  xleftmargin=2em, xrightmargin=0em, 
}

\newcommand\nmg[1]{{\color{black}{#1}}}

\newcommand\klgh[1]{{\color{black}{#1}}}

\usepackage{enumitem}

\makeatletter
\g@addto@macro{\normalsize}{%
  \setlength{\abovedisplayskip}{4pt plus 0.5pt minus 1pt}
  \setlength{\belowdisplayskip}{4pt plus 0.5pt minus 1pt}
  \setlength{\abovedisplayshortskip}{0pt}
  \setlength{\belowdisplayshortskip}{0pt}
  \setlength{\intextsep}{3pt plus 1pt minus 1pt}
  \setlength{\textfloatsep}{7pt plus 1pt minus 1pt}
  \setlength{\skip\footins}{5pt plus 1pt minus 0pt}}
\makeatother

\presetkeys{todonotes}{linecolor=black,backgroundcolor=white,bordercolor=black}{}

\renewcommand\revmid[2]{\todo[linecolor=#1,backgroundcolor=white,bordercolor=#1]{\textcolor{#1}{\textbf{#2}}}}

\hypersetup{colorlinks=false,
            citebordercolor={0 1 0},   %
            linkbordercolor={1 0 0},   %
            urlbordercolor={0 1 1}}    %

\providecommand{\circled}[1]{\tikz[baseline=(char.base)]{
    \node[shape=circle,draw,inner sep=1pt] (char) {#1};}}

\newcommand{\param}[1]{\textcolor{black}{#1}} %
\providecommand{\head}[1]{\noindent\textbf{\\#1}\\} %
\newcommand{\xxx}[1]{\param{XXX}} %

\newcommand{\ignore}[1]{}

\newcounter{obs}
\newcounter{tkw}
\definecolor{nbs}{rgb}{0.88, 0.07, 0.37}
\definecolor{iy}{rgb}{0.0, 0.56, 0.041}
\definecolor{akkpink}{rgb}{0.88, 0.28, 0.54}
\definecolor{hscolor}{rgb}{0.529, 0.298, 0.384}

\newif\ifdraft
\draftfalse

\ifdraft
    \usepackage[colorinlistoftodos,prependcaption,textsize=small]{todonotes}
    \paperwidth=\dimexpr \paperwidth + 4cm\relax
    \oddsidemargin=\dimexpr\oddsidemargin + 2cm\relax
    \evensidemargin=\dimexpr\evensidemargin + 2cm\relax
    \marginparwidth=\dimexpr \marginparwidth + 2cm\relax

    \newcommand{\nbcomment}[1]{\todo[size=\scriptsize, linecolor=orange, bordercolor=orange, backgroundcolor=white]{\textcolor{nbs}{\textbf{@nb:} #1}}}

    \newcommand{\ieycomment}[1]{\todo[size=\scriptsize, linecolor=orange, bordercolor=orange, backgroundcolor=white]{\textcolor{iy}{\textbf{@iey:} #1}}}

    \newcommand{\atbcomment}[1]{\todo[size=\scriptsize, linecolor=orange, bordercolor=orange, backgroundcolor=white]{\textcolor{blue}{\textbf{@atb:} #1}}}
    
    \newcommand{\dccomment}[1]{\todo[size=\scriptsize, linecolor=orange, bordercolor=orange, backgroundcolor=white]{\textcolor{violet}{\textbf{@dc:} #1}}}
    \newcommand{\hscomment}[1]{\todo[size=\scriptsize, linecolor=orange, bordercolor=orange, backgroundcolor=white]{\textcolor{hscolor}{\textbf{@hs:} #1}}}

\else 
    \newcommand{\nbcomment}[1]{}

    \newcommand{\ieycomment}[1]{}

    \newcommand\atbcomment[1]{}
    
    \newcommand\dccomment[1]{}
    \newcommand\hscomment[1]{}
    
\fi

\newcommand{\secref}[1]{§\ref{#1}}

\newif\ifrevision
\revisiontrue

\definecolor{nbs}{rgb}{0.88, 0.07, 0.37}
\definecolor{iy}{rgb}{0.0, 0.36, 0.05}
\definecolor{gfored}{rgb}{0.580, 0.050, 0.211}
\definecolor{bluepigment}{rgb}{0.2, 0.2, 0.6}
\definecolor{brandeisblue}{rgb}{0.0, 0.44, 1.0}
\definecolor{ouscolor}{rgb}{0.0, 0.2, 0.4}
\definecolor{omc}{rgb}{1, 0.01, 0.01}

\definecolor{ra}{HTML}{E6194B}       %
\definecolor{rb}{HTML}{4363D8}       %
\definecolor{rc}{HTML}{3CB44B}       %
\definecolor{rd}{HTML}{E76F51}
\definecolor{re}{HTML}{808000}       %
\definecolor{rf}{HTML}{911EB4}       %

\definecolor{cq}{HTML}{42D4F4}       %
\definecolor{cqtwo}{HTML}{F032E6}    %
\definecolor{cqthree}{HTML}{9A6324}  %
\definecolor{cqfour}{HTML}{F58231}   %
\definecolor{cqfive}{HTML}{469990}   %

\definecolor{cq}{rgb}{0.0, 0.5, 1.0}

\ifrevision
    
    \newcommand{\reva}[1]{\textcolor{ra}{#1}}
    \newcommand{\revb}[1]{\textcolor{rb}{#1}}
    
    \newcommand{\revd}[1]{\textcolor{rd}{#1}}
    \newcommand{\reve}[1]{\textcolor{re}{#1}}
    
    \newcommand{\revcqone}[1]{\textcolor{cq}{#1}}
    \newcommand{\revcqtwo}[1]{\textcolor{cqtwo}{#1}}
    \newcommand{\revcqthree}[1]{\textcolor{cqthree}{#1}}
    \newcommand{\revcqfour}[1]{\textcolor{cqfour}{#1}}

    \definecolor{babyblueeyes}{rgb}{0.63, 0.79, 0.95} 
    
\else
    
    \newcommand{\reva}[1]{{#1}}
    \newcommand{\revb}[1]{{#1}}
    
    \newcommand{\revd}[1]{{#1}}
    \newcommand{\reve}[1]{{#1}}

    \providecommand\revmid[2]{}

    \definecolor{babyblueeyes}{rgb}{0.63, 0.79, 0.95} 
\fi

\ifrevision
    
\else
    
\fi

\newif\ifcleanmode
\cleanmodetrue
\ifcleanmode
  \def\reva#1{#1}\def\revb#1{#1}\def\revd#1{#1}\def\reve#1{#1}

  \def\revcqone#1{#1}\def\revcqtwo#1{#1}\def\revcqthree#1{#1}\def\revcqfour#1{#1}

  \def\revmid#1#2{}
  \AtBeginDocument{\hypersetup{colorlinks=true,allcolors=black}}
\fi

\ifarxiv
  \renewcommand{\labela}[1]{}\renewcommand{\labelb}[1]{}\renewcommand{\labelc}[1]{}
  \renewcommand{\labeld}[1]{}\renewcommand{\labele}[1]{}\renewcommand{\labelf}[1]{}
  \renewcommand{\labelcq}[1]{}\renewcommand{\labelcqtwo}[1]{}\renewcommand{\labelcqthree}[1]{}
  \renewcommand{\labelcqfour}[1]{}\renewcommand{\labelcqfive}[1]{}
\fi

\newcommand{\citepersonalized}{alkan_personalized_2009,lightbody_review_2019,morganti_next_2019,branco_bioinformatics_2021,quazi_artificial_2022,aronson_building_2015,f_lochel_comparative_2020,papadopoulou_application_2023,tafazoli_applying_2021,gambardella_personalized_2020,leary_development_2010,hamburg_margaret_a_path_2010,van_der_lee_technologies_2020,moon_precision_2022,mohan_profiling_2020,chung_rapid_2020,bielinski_preemptive_2014,ho_enabling_2020,hussen_emerging_2022,russell_pharmacogenomics_2021,verma_nanopore_2024,clark2019diagnosis,farnaes2018rapid,sweeney2021rapid,flores2013p4,ginsburg2009genomic,chin2011cancer,Ashley2016}

\newcommand{\citeoutbreakrapid}{dunn_squigglefilter_2021,bertelli_rapid_2013,arias_rapid_2016,comin_investigation_2020,Quick2016}

\newcommand{\citeoutbreak}{\citeoutbreakrapid,robinson_genomics_2013,fournier_clinical_2014,koser_routine_2012,eloit_diagnosis_2014,gardy_jennifer_l_whole-genome_2011,taylor_angela_j_characterization_2015,quainoo_scott_whole-genome_2017,goldberg_brittany_making_2015,besser_interpretation_2019,li_application_2021,deng_integrated_2021,kwong_whole_2015,deurenberg_application_2017,tang_infection_2017,croucher_application_2015,
bloom2021massively,yelagandula2021multiplexed,le2013selected,nikolayevskyy2016whole,qiu2015whole,gilchrist2015whole}

\newcommand{\citeagriculture}{the_arabidopsis_genome_initiative_analysis_2000,zhu_applications_2020,choi_nanopore_2020,stevens_sequence_2016,campos_high_2021,gao_genome_2021,van_dijk_machine_2021,sun_twenty_2022,kim_application_2020,thudi_genomic_2021,michael_building_2020,shen_omics-based_2022,shahroodi2022demeter,
prasad2021soil,Mascher2024,Schreiber2024}

\newcommand{\citeevolution}{kanehisa_toward_2019,qing_whole_2022,wittkopp_cis-regulatory_2012,romero_comparative_2012,wang_population_2020,hill_molecular_2021,vaishnav_evolution_2022,zhang_haplotype-resolved_2021,fay_evaluating_2008,kanzi_next_2020,wray_evolution_2003,wu_one_2021,signor_evolution_2018,whitehead_variation_2006,coolon_tempo_2014,
ellegren2014genome,Prado-Martinez2013,Prohaska2019human}

\newcommand{\citecancer}{lawrence_mutational_2013,vogelstein_cancer_2013,ramskold_full-length_2012,baslan_unravelling_2017,shapiro_single-cell_2013,sakamoto_new_2020,jia_high-throughput_2022,lawson_tumour_2018,liu_mrna-based_2023,van_de_sande_applications_2023,chakravarty_clinical_2021,cortes-ciriano_computational_2022,deveson_evaluating_2021,xiao_toward_2021,bolton_cancer_2020,szustakowski_advancing_2021,navin_future_2011,hong_rna_2020,lei_applications_2021,han_single-cell_2022,federici_variants_2020,zhang_singlecell_2021,ren_understanding_2018,tian_cicero_2020,malone_molecular_2020,tang_single-cell_2019,ellsworth_single-cell_2017,zhong_application_2021,stadler_therapeutic_2021,tan_targeted_2022,degasperi_substitution_2022,xu_single-cell_2022,horak_comprehensive_2021,zhang_single-cell_2016,bruno_next_2020,de_luca_fgfr_2020,waarts_targeting_2022,lim_advancing_2020,colomer_when_2020,saadatpour_single-cell_2015,dizman_sequencing_2020,buzdin_rna_2020,xiao_tumor_2021,nandwani_lncrnas_2021,marchetti_error-corrected_2023,chen_next-generation_2021,navin_first_2015}

\newcommand{\citehwoptimization}{doblas2025smx,mutlu2023accelerating,alser2022molecules,lou2020helix,lou2018brawl,shahroodi2023swordfish,markus2020benchmarking,subramaniyan2021accelerated,huangfu2018radar,khatamifard2021genvom,gupta2019rapid,li2021pim,angizi2019aligns,zokaee2018aligner,turakhia2018darwin,fujiki2018genax,madhavan2014race,cheng2018bitmapper2,houtgast2018hardware,houtgast2017efficient,zeni2020logan,ahmed2019gasal2,nishimura2017accelerating,de2016cudalign,liu2015gswabe,liu2013cudasw++,liu2009cudasw++,liu2010cudasw++,wilton2015arioc,goyal2017ultra,chen2016spark,chen2014accelerating,chen2021high,fujiki2020seedex,banerjee2018asap,fei2018fpgasw,waidyasooriya2015hardware,chen2015novel,rucci2018swifold,haghi2021fpga,li2021pipebsw,ham2020genesis,ham2021accelerating,wu2019fpga,cali2020genasm,Zhang_2023_alignerD,soysal2025mars,kim2018grim,kaplan2020bioseal,mao2022genpip,dphls2026,wang20202,Walia2024talco,sadasivan2024genomic,Turakhia2025toward,Turakhia2019darwinwga,simon2026pim}
\newcommand{\citemghwoptimization}{jia2011metabing,kobus2021metacache,wang2023gpmeta,kobus2017accelerating,Su2012,su2013gpumetastorms,Yano2014,saavedra2020mining,zhang2023genomix,cervi2022metagenomic,wu2021sieve,shahroodi2022krakenonmem,shahroodi2022demeter,dashcam23micro,hanhan2022edam,zou2022biohd,dunn_squigglefilter_2021,shih2023efficient}

\newcommand{\citegraphhwoptimization}{Cali2022SeGraM,Zhang2024,Zeng2024asgdp,Shen2024128parallel,Li2024,Mandal2020,Varma2013,Awan2021,Feng2021,Zhang2025,kim2025nmp,Huang2023meg2,Angizi2020Panda,Qiu2017,Zhou2021,Sarkar2021,Varma2017,Varma2016,Goswami2018,Galanos2021,Angizi2020,Sinha2022,Meng2014,Hu2016,Chen2023,Natarajan2018,Ren2018}

\hypersetup{pdftitle={Motivo: A Motif-Based Algorithm-Architecture Co-Design for Fast and Energy-Efficient Graph-Based Genome Analysis},
  pdfauthor={Eirini Maria Tzermpou, Konstantina Koliogeorgi, Konstantinos Kanellopoulos, Ataberk Olgun, Nisa Bostanc{\i}, Dimitrios Soudris, Nika Mansouri Ghiasi, Onur Mutlu}}

\begin{document}

\author{Eirini Maria Tzermpou$^{1}$
\quad Konstantina Koliogeorgi$^{1}$
\quad Konstantinos Kanellopoulos$^{1}$
\quad Ataberk Olgun$^{1}$ \\
Nisa Bostanc{\i}$^{1}$
\quad Dimitrios Soudris$^{2}$
\quad Nika Mansouri Ghiasi$^{1}$
\quad Onur Mutlu$^{1}$
\vspace{-4mm} \\\\
$^{1}$ETH Z{\"u}rich \qquad $^{2}$NTUA}

\fancyhead{}
\renewcommand{\headrulewidth}{0pt}

\title{Motivo: A Motif-Based Algorithm-Architecture Co-Design\\for Fast and Energy-Efficient Graph-Based Genome Analysis}

\maketitle
\setstretch{0.97}
\begin{abstract}
\def\UseAbsKonkanello{}\providecolor{darkgreen}{rgb}{0.0, 0.39, 0.0}%
\providecommand\dg[1]{{\color{black}{#1}}}%
\providecommand\tgmasic{\textsf{\toolname{}-ASIC}\xspace}%
\providecommand\tgmcpu{\textsf{\toolname{}-CPU}\xspace}%
\ifdefined\UseAbsKonkanello%
Genome sequence analysis, which examines the genomic information of an organism, plays a critical role in many fields. \textit{Read mapping}, a key step of genome sequence analysis, matches each fragment of a sequenced genome, called a \textit{read}, to its location in a known \emph{reference genome}. A \emph{genome graph} is a graph-based reference genome that represents the genetic diversity of a population as alternative paths, and thus improves read mapping accuracy. However, the size and complexity of a genome graph make sequence-to-graph (S2G) mapping, i.e., mapping a read to a genome graph, a major bottleneck of genome sequence analysis. We observe that prior software and hardware approaches for S2G mapping perform \emph{redundant} costly operations (e.g., graph traversals and dynamic programming) that do \emph{not} lead to a valid mapping.

Our goal in this work is to deliver a high-performance and energy-efficient design for S2G mapping by reducing redundant computation and replacing expensive computation with lightweight operations. \dg{Our key observation is that a genome graph consists of small subgraphs that share a limited set of recurring topologies, which we call \emph{motifs}. Based on this observation, we propose \tgm{}, a topology-aware algorithm-architecture co-design for S2G mapping. At the core of \tgm{} is a new motif-based S2G mapping algorithm that replaces graph traversals and dynamic programming on every candidate path with memory lookups and arithmetic operations.
The algorithm} uses motifs in two ways. First, \dg{it} stores the precomputed motifs of each node in a compact index, which replaces a graph traversal with a single memory lookup. Second, \dg{it} uses the topology of a motif and the properties of its nodes (i.e., sequence length and base composition) to identify the few most suitable mapping paths with arithmetic operations.
\dg{\tgm{}'s algorithm enables \inum{i}~high-performance execution on two conventional hardware platforms (\tgmcpu{} and \tgmgpu{}) and \inum{ii}~high-performance and low-energy execution on \tgmasic{}, a new ASIC-based accelerator that we co-design with the algorithm. Since the path selection of the algorithm uses only memory lookups and arithmetic operations, \tgmasic{} consists of simple hardware units.}

\dg{\tgmcpu{} provides 2.1-18.4$\times$, 2.8-343$\times$, and 1.4-10.1$\times$ speedups over GraphAligner, vg, and minigraph, respectively, and \tgmgpu{} provides 1.5-1.6$\times$ speedups over vg-giraffe. \tgmasic{} provides 7.4-11.7$\times$ (138-274$\times$) speedups over the state-of-the-art CPU-based tools for short (long) reads.} Compared to two state-of-the-art ASIC-based accelerators, \dg{SeGraM (short reads) and HARP (long reads), \tgmasic{} achieves \textcolor{black}{5.3$\times$} and \textcolor{black}{4.0-4.3$\times$} higher throughput per Watt and \textcolor{black}{1.7$\times$} and \textcolor{black}{2.7-3.0$\times$} higher throughput per mm$^2$, respectively, and \textcolor{black}{1.9-2.7$\times$} higher throughput than SeGraM}. \tgm{} maintains equal or higher mapping accuracy compared to other tools \textcolor{black}{(96.8\%-99.7\%)} \dg{for short and long reads on two types of genome graphs}.
\fi

\end{abstract}
\setstretch{1}

\providecolor{darkgreen}{rgb}{0.0, 0.39, 0.0}
\providecommand\dg[1]{{\color{black}{#1}}}
\providecommand\tgmasic{\textsf{\toolname{}-ASIC}\xspace}
\providecommand\tgmcpu{\textsf{\toolname{}-CPU}\xspace}
\section{Introduction}
\label{sec:introduction}

Genome sequence analysis, which \dg{examines the genomic information of an organism}, plays a critical role in many fields, such as personalized medicine~\cite{\citepersonalized}, cancer research~\cite{\citecancer}, disease surveillance~\cite{\citeoutbreak}, evolutionary biology~\cite{\citeevolution}, and agriculture~\cite{\citeagriculture}.
\dg{A sequencing machine produces many fragments of the DNA of an organism, called \textit{reads}}~\cite{sanger1977dna}. \dg{A read is a \textit{short read} (up to a few hundred bases) or a \textit{long read} (thousands to millions of bases).}
\dg{To keep up with the growing amount of sequencing data, many prior works accelerate genome sequence analysis} (e.g.,~\cite{\citehwoptimization,\citemghwoptimization,\citegraphhwoptimization,singh2024rubicon,cavlak_targetcall_2024,Simon2025cimba,Firtina2024aphmm,firtina2020apollo,megis}).
\dg{\emph{Read mapping}, a key step of genome sequence analysis, matches each read to its location in a known \textit{reference genome} (i.e., a sequence that is representative of a species). When the reference genome is a single sequence, read mapping is a \textit{sequence-to-sequence (S2S) mapping} problem.}

\begin{figure}[h]
    \centering
    \includegraphics[width=0.9\columnwidth]{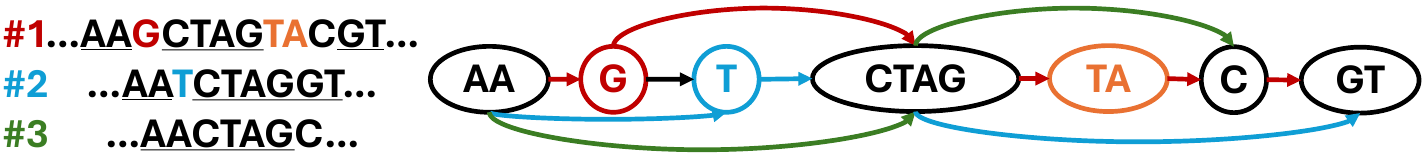}
    \caption{\textcolor{black}{A genome graph built from three genomes (\#1-\#3). Each node stores a sequence, and each color marks the path of one genome through the graph.}}
    \label{fig:ref_gr_constr}
\end{figure}

\dg{A single reference genome does not represent the genetic variation across the individuals of a species, which reduces the mapping accuracy for a read that differs from the reference genome~\cite{paten2017genome,Sherman2020,garrison2018variation}.
A \emph{genome graph} is a graph-based reference genome that represents the genetic diversity of a population as alternative paths, and thus improves read mapping accuracy}~\cite{eizenga2020pangenome,taylor2024beyond,Liao2023,Sherman2020,Mustafa2024MLA}.
\dg{As \fig{\ref{fig:ref_gr_constr}} shows, each node of a genome graph stores a sequence, each edge connects two nodes whose sequences can follow each other, and each path through the graph spells a possible genome sequence.}
\dg{A \textit{variant}, i.e., a difference between two genomes, forms alternative paths in the graph.}
\dg{Genome graphs enable new directions in applications such as} population-scale~\cite{eizenga2020pangenome,Armstrong2020cactus,bradley2019ultrafast,Sherman2020} and personalized genomics~\cite{\citepersonalized}.

\dg{Mapping a read to a genome graph is a \textit{sequence-to-graph (S2G) mapping} problem: an S2G mapping tool finds the path of the genome graph whose sequence is the most similar to the read}~\cite{Jain_2020,rautiainen2020graphaligner,Hickey2020,Cui2025}. \dg{Before mapping, the tool builds a \textit{graph index} of the genome graph, which enables fast lookup of sequences in the graph}~\cite{Siren_2017,Cui2025}. S2G mapping typically follows a \textit{seed-and-extend} strategy \dg{with} three steps: \textit{seeding}, \textit{filtering}, and \textit{extension}. First, \dg{the \textit{seeding} step extracts short subsequences of the read (\textit{seeds}) and queries the graph index to find the positions in the graph that each seed matches}, i.e., \textit{seed hits}~\cite{Cui2025}. Second, the \textit{filtering} step discards, clusters, or chains \dg{the} seed hits \dg{to reduce the number of candidates for the extension step} (e.g.,~\cite{siren2021pangenomics,rautiainen2020graphaligner,Li2020minigraph}).
Third, the \textit{extension} step expands each \dg{remaining} seed hit into a local \textit{seed hit subgraph} and aligns the read against each path in it, producing a \textit{candidate alignment} per path. \textit{Alignment} compares two sequences base by base and finds the fewest changes (i.e., \textit{edits}) that turn one into the other~\cite{lee2002multiple,marco2021fast,alser2020accelerating}. An alignment is \dg{\textit{valid}} if \dg{its} number of edits does not exceed a threshold.
Alignment \nmg{is} the computational bottleneck of both S2S and S2G \dg{mapping}, as it relies on expensive dynamic programming (DP) to \dg{find the edits} between a read and a reference sequence~\cite{alser2017gatekeeper,fujiki2018genax,ham2020genesis,Huangfu2019medal,nag2019gencache,turakhia2018darwin}.

Despite its accuracy benefits, S2G mapping poses \dg{two key} challenges. First, \dg{a genome graph has many alternative paths, which lead to more candidate alignments than the single sequence of a reference genome}~\cite{Jain_2020,jain2019pasgal,garrison2018variation}. \dg{More candidate alignments increase the cost of the} already expensive alignment step~\cite{needleman1970general,xia2022review,bingol2026debruijn,PangenomicsBench}.
Second, \dg{a genome graph is large: it contains} hundreds of millions of nodes~\cite{Li2020minigraph,Liao2023}. \dg{As sequencing technologies produce more data at lower cost, the number of genomes that a genome graph represents grows, and so does the graph. In a larger graph, a read has more seed hits, each with many alternative paths, which further increases the cost of S2G mapping~\cite{bingol2026debruijn,Cui2025}.}

To alleviate these challenges, software tools optimize the individual steps of \dg{S2G mapping}, such as \dg{indexing}~\cite{Altschul_1990,Langmead_2009,burrows1994block,huang2013short,siren2014indexing,ferragina2000opportunistic,rossi2022moni,siren2020haplotype}, \dg{seeding}~\cite{schneeberger2009simultaneous,heydari2018browniealigner,ma2023chaining,sahlin2022strobealign,li2012exploring}, filtering~\cite{limasset2016read,dvorkina2020spaligner,garrison2018variation,rautiainen2020graphaligner,siren2021pangenomics,Li2020minigraph,liu2016debga,li2018minimap2,chandra2023sequence}, and extension~\cite{marco2021fast,zhang2022fast,Ivanov2020,ivanov2022fast,needleman1970general,xia2022review,lee2002multiple,monti2026faster}. \dg{Other works speed up S2G mapping with graphics processing units (GPUs)}~\cite{Feng2021,zhu2025parabricks} or \dg{with hardware accelerators based on application-specific integrated circuits (ASICs)}~\cite{Cali2022SeGraM,Zhang2024}.

\noindent\textbf{Key \dg{Limitation}.} \dg{Despite their benefits, both software tools and hardware accelerators for S2G mapping perform \textit{redundant} operations, i.e., costly operations (e.g., graph traversals and DP) on candidates that do \emph{not} lead to a valid alignment.
Our analysis of state-of-the-art S2G mapping tools (\sect{\ref{sec:motivation}}) shows that operations on discarded candidates alone account for 31-99.3\% of the end-to-end mapping time for short reads and 83-99.9\% for long reads.
Existing filtering steps cannot eliminate these redundant operations.}

\begin{figure}[h]
    \centering
    \includegraphics[width=0.9\columnwidth]{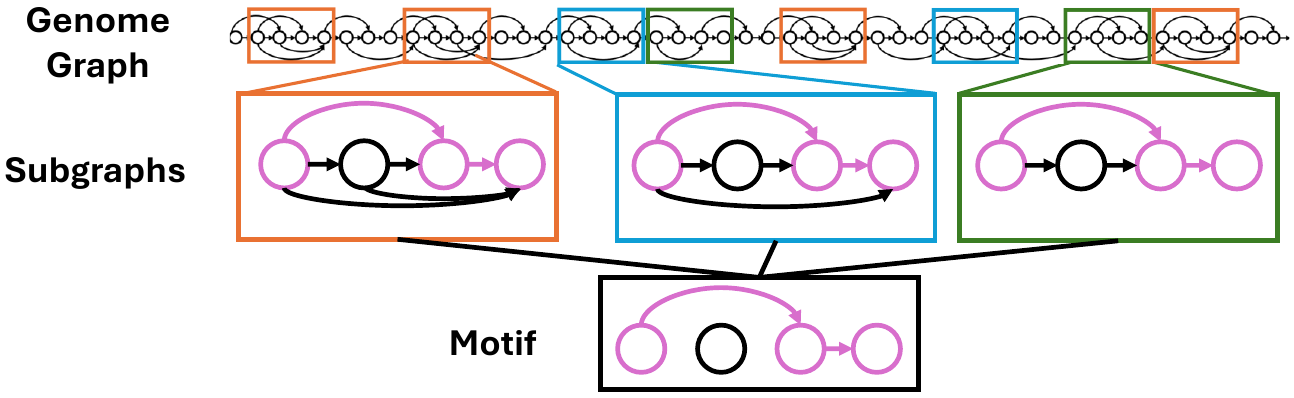}
    \caption{\textcolor{black}{Three subgraphs of a genome graph (orange, blue, and green boxes) contain the same motif (pink): a path over four consecutive nodes that skips the second node.}}
    \label{fig:gr_subg_motif}
\end{figure}

\textbf{Our goal} in this work is to deliver a high-performance and energy-efficient design for S2G mapping by reducing redundant computation and replacing expensive computation with lightweight operations.
\dg{Our \textbf{key observation} is that a genome graph consists of small subgraphs that share a limited set of recurring topologies, which we call \emph{motifs}. Most genetic variation forms a \textit{small hop}, i.e., a short branch that leaves and rejoins the rest of the graph within a few nodes (\sect{\ref{sec:motivation}}), so the same small topologies recur across the genome graph (\fig{\ref{fig:gr_subg_motif}}).}

\dg{Based on this observation,} we propose \tgm{}, a \klgh{topology-aware algorithm-architecture co-design for sequence-to-graph (S2G) mapping}.
\dg{At the core of \tgm{} is a new motif-based S2G mapping algorithm that replaces graph traversals and DP on every candidate path with memory lookups and arithmetic operations (e.g., additions and comparisons). The \textbf{key idea} of the algorithm is to use motifs to identify, \emph{before} alignment, the few paths around a seed hit that can lead to a valid alignment. The algorithm aligns the read only to these paths. \tgm{}'s algorithm enables \inum{i}~high-performance execution on two conventional hardware platforms, i.e., a CPU and a GPU (\tgmcpu{} and \tgmgpu{}), and \inum{ii}~high-performance and low-energy execution on \tgmasic{}, a new ASIC-based accelerator that we co-design with the algorithm.}

\noindent\textbf{Key Mechanism.} \dg{\tgm{} has three key components (\sect{\ref{sec:overview}}).
First, \tgmindex{} is a compact graph index that stores, for each node of the genome graph, its precomputed motifs and its properties (i.e., sequence length and base composition).
Second, \tgmext{} is a new extension mechanism that finds the} \revcqone{most suitable mapping paths} \dg{for each seed hit with memory lookups and arithmetic operations. \tgmext{} retrieves the motifs of the seed hit from \tgmindex{}, discards the candidate paths that cannot lead to a valid alignment, and} \revcqone{ranks the remaining ones to keep only the top-$K$ paths}\dg{. As a result, \tgm{} reduces S2G mapping to a few S2S alignments per seed hit.
Third, we implement the algorithm of \tgm{} on three hardware platforms. \tgmcpu{} and \tgmgpu{} are straightforward software implementations of the algorithm on a CPU and on a GPU. They outperform highly optimized state-of-the-art CPU-based and GPU-based tools, which shows that the benefits of \tgm{} come from the algorithm itself. \tgmasic{} executes the algorithm end to end in hardware. Since the path selection of the algorithm uses only memory lookups and arithmetic operations, \tgmasic{} consists of simple hardware units and provides \textcolor{black}{high throughput per unit of area and power}.}

\noindent\textbf{Key Results.} \revcqfour{We evaluate \tgm{} with short and long reads} \revcqtwo{on \dg{two types of genome graphs:} the \dg{Genome in a Bottle (GIAB)} graphs used by prior work~\cite{Cali2022SeGraM} and the denser \dg{Human Pangenome Reference Consortium (HPRC)} graphs~\cite{Liao2023}}.
We compare \tgm{} against state-of-the-art \dg{CPU-based} (GraphAligner~\cite{rautiainen2020graphaligner}\revcqfour{, vg~\cite{garrison2018variation}, minigraph~\cite{Li2020minigraph}}), \dg{GPU-based} (vg-giraffe~\cite{nvidia_parabricks_giraffe}), and \dg{ASIC-based} (SeGraM~\cite{Cali2022SeGraM}\revcqfour{, HARP~\cite{Zhang2024}}) systems. We make \dg{four} key observations.
First, \dg{the algorithm of \tgm{} alone improves performance. \tgmcpu{} provides 2.1-18.4$\times$, 2.8-343$\times$, and 1.4-10.1$\times$ speedups over GraphAligner, vg, and minigraph, respectively, and \tgmgpu{} provides 1.5-1.6$\times$ speedups over vg-giraffe.}
Second, \dg{\tgmasic{} provides 7.4-11.7$\times$ and 7.7-8.2$\times$ speedups over the state-of-the-art CPU- and GPU-based baselines for short reads, respectively, and 138-274$\times$ speedups over the state-of-the-art CPU-based baselines for long reads. With the same algorithm, \tgmasic{} is 4.2-6.9$\times$ (68-113$\times$) faster than \tgmcpu{} for short (long) reads, and 5.2-5.3$\times$ faster than \tgmgpu{}.}
Third, \dg{\tgmasic{} outperforms two state-of-the-art ASIC-based accelerators. For short reads on the GIAB graphs, \tgmasic{} provides} \textcolor{black}{1.9-2.7$\times$} \dg{higher throughput,} \textcolor{black}{5.3$\times$} higher throughput per Watt\dg{, and \textcolor{black}{1.7$\times$} higher throughput per mm$^2$ than SeGraM. For long reads on the HPRC graphs, \tgmasic{} provides \textcolor{black}{4.0-4.3$\times$} higher throughput per Watt and \textcolor{black}{2.7-3.0$\times$} higher throughput per mm$^2$ than HARP.}
\dg{Fourth, \tgm{} achieves \textcolor{black}{99.1\%} (\textcolor{black}{99.7\%}) mapping accuracy for short (long) reads on the GIAB graphs and \textcolor{black}{98.0\%} (\textcolor{black}{96.8\%}) on the HPRC graphs, which is even higher than some state-of-the-art baselines.}

This work makes the following key contributions:
\begin{itemize}[leftmargin=*, topsep=0pt, partopsep=0pt, itemsep=2pt, parsep=0pt]

\item We propose \textbf{\tgm{}}\dg{.} \revb{\labelb{5}\dg{To our knowledge, \tgm{} is the first algorithm-architecture co-design for sequence-to-graph (S2G) mapping that exploits recurring subgraph topologies.}} \dg{\tgm{} uses motifs to identify, \emph{before} alignment, the few paths of a genome graph that can lead to a valid alignment of a read.}

\item \dg{We introduce two key mechanisms of \tgm{}: \inum{i} \tgmindex{}, a motif-based graph index that stores the motifs and the properties of each node of a genome graph, and \inum{ii} \tgmext{}, a mechanism that uses \tgmindex{} to identify the} \revcqone{\dg{few most suitable mapping paths}} \dg{for a read with lightweight arithmetic operations, instead of dynamic programming on every candidate path.}

\item \dg{We show that the algorithm of \tgm{} improves the performance of S2G mapping on a CPU and on a GPU, and we co-design the algorithm with \tgmasic{}, a new hardware accelerator based on an application-specific integrated circuit (ASIC).}

\item We show that \tgm{} \dg{provides higher throughput and energy efficiency than} state-of-the-art \dg{CPU-based, GPU-based, and ASIC-based} S2G mapping systems \revcqfour{for both short and long reads}. Compared to \dg{two} state-of-the-art \dg{ASIC-based} accelerators, \dg{\tgmasic{} provides up to 2.7$\times$ higher throughput, 5.3$\times$ higher throughput per Watt, and 3.0$\times$ higher throughput per mm$^2$, with {96.8\%-99.7\%} mapping accuracy.}

\end{itemize}

\section{Background}
\label{sec:background}
\subsection{Genome Sequence Analysis}
A genomic sample is processed in three main steps: \inum{i} sequencing, \inum{ii} basecalling, and \inum{iii} analysis. The first step, sequencing, converts the sample's genomic information (e.g., \nmg{encoded in} DNA or RNA, which is typically converted to DNA before sequencing~\cite{Houldcroft2017,Jansz2024viral}) into digital signals. Sequencing machines cannot provide the end-to-end sequence of an entire DNA molecule, and instead, they break it down into many short, overlapping fragments called \emph{reads}. Basecalling then converts the raw signal of each read into a string of four nucleotide bases: A, C, G, T~\cite{cock2009sanger}. Read length (i.e., the number of bases of a read) is measured using base pairs (bp). \emph{Genome analysis} takes the resulting set of strings, i.e., a \emph{read set}, and compares it against a \emph{reference genome}, i.e., a known sequence that represents a species, or against large-scale databases that contain reference genomes of many species and previously sequenced samples. 
A major step of genome analysis is \emph{read mapping}, where each read of a sequenced sample is
matched against the reference genome of the organism's species to find the read's original location and the differences between the two~\cite{alser2020accelerating,li2018minimap2}. \emph{Alignment} identifies these differences by comparing the two sequences base by base. It computes their \emph{edit distance}, i.e., the minimum number of base substitutions, insertions, and deletions needed to transform one sequence into the other~\cite{smith1981identification}.
\subsection{Genome Graph Representations}

Linear reference representations of genomes, i.e., genome representations based on a single or a few individuals, fail to represent population-level genetic diversity~\cite{Nurk_2022,paten2017genome}. Individual genomes differ from a reference through several classes of variation. The most common is the \textit{single-nucleotide polymorphism} (SNP): a substitution of one base for another at a single position (e.g., an A where the reference has a G). \textit{Insertions} and \textit{deletions} (\textit{indels}) add or remove one or more bases, while \textit{structural variants} (e.g., large indels, duplications, and inversions) involve longer stretches of sequence~\cite{Hickey2020}. 
Since a linear reference genome can represent only one of the many sequence variants found in a population at each position, researchers have shifted towards graph representations~\cite{Cali2022SeGraM,marchet2021data,karasikov2020metagraph,fan2023fulgor,bradley2019ultrafast,Muggli2017SuccinctGraphs,Iqbal2012,turner2018integrating,hunt2024allthebacteria}, which integrate information from multiple genomes and accurately capture a population's genetic diversity.
In a genome graph, \textit{nodes} correspond to genomic sequences, \textit{edges} denote adjacency relationships between them, and individual \textit{paths} correspond to single genomes~\cite{garrison2018variation,Li2020minigraph}. Among the graph models that have been proposed, one of the most commonly used types is \textbf{variation graphs}, which hold all variation information in a single structure. 

Shared sequence among different genomes is consolidated into common nodes, while variation appears as divergent structure: an SNP forms a short branch where alternative single-base paths rejoin after one node, whereas indels and structural variants create longer paths that diverge from and later rejoin the main sequence~\cite{Hickey2020}. 
This yields a compact encoding of the shared backbone while still expressing complex, individual-specific variation~\cite{siren2020haplotype}.

\subsection{S2G Mapping Workflow Overview}
\label{sec:back-e}

In traditional linear reference mapping (S2S), the goal is to identify the most suitable location within a reference string that matches the query sequence while minimizing the edit distance~\cite{smith1981identification,li2018minimap2}.
In contrast, sequence-to-graph (S2G) mapping generalizes this problem to graph-based references, where the goal shifts to identifying the most probable alignment walk through the graph~\cite{Jain_2020,garrison2018variation,rautiainen2020graphaligner}.
This requires jointly optimizing path selection and alignment, substantially increasing complexity but enabling more accurate mapping in regions of high variation~\cite{siren2021pangenomics}.
Most S2G mapping algorithms include three online steps: \textit{seeding}, \textit{clustering/filtering}, and \textit{extension}~\cite{garrison2018variation,siren2021pangenomics,rautiainen2020graphaligner,Li2020minigraph,dvorkina2020spaligner,jain2019pasgal,Ivanov2020,ma2023chaining,chandra2023sequence}, while they use an offline pre-built reference graph index for all queries.

\noindent\textbf{Offline Reference Graph Indexing.} To enable fast lookup of short matches during mapping, an index is built over the graph's sequences before mapping (i.e., offline). Common designs include hash tables, compressed full-text indexes such as FM-/BWT-based structures~\cite{li_bwt_2024}, and lightweight \textit{sketch} indexes. Sketching selects a small subset of a sequence's \textit{k-mers} (i.e., its subsequences of length $k$) according to a fixed rule, so that only the selected k-mers are stored instead of all of them~\cite{siren2014indexing,Siren_2017,roberts2004reducing}. A widely used, memory-efficient sketching rule produces \textit{minimizers}: given a k-mer length $k$ and a window size $w$, a windowed scheme slides across the sequence and, within each window of consecutive k-mers, selects the k-mer with the smallest hash value as the minimizer~\cite{roberts2004reducing}. The index stores each minimizer's hash together with its position in the graph (i.e., the node ID and the start offset within that node's sequence).
\noindent\textbf{Seeding.} The first stage of S2G mapping (i.e., seeding) finds short exact or near-exact matches between the read and the graph's node sequences~\cite{Cui2025}. Short subsequences are extracted from the read by applying the same sketching rule used during index construction (e.g., minimizers), and are then queried against the index to retrieve their positions in the graph~\cite{roberts2004reducing,li2016minimap,Li_2009,Jung_2022,Siren_2017,garrison2018variation,rautiainen2020graphaligner,Li2020minigraph,Jain_2020,Fernandes_2013,li2018minimap2,Edgar_2021,Sahlin_2021,sahlin2022strobealign}.

\noindent\textbf{Clustering/Filtering/Chaining.} A single read can produce many seed hits across the graph, since short subsequences appear in many places, but only a few of them lead to valid alignments. This step reduces the number of seed hits that the mapper needs to examine~\cite{Cui2025}. \textit{Filtering} discards the seed hits that are unlikely to lead to a valid alignment, e.g., those that are isolated in the graph. \textit{Clustering} groups together seed hits that are found close to each other in the graph, since they likely point to the same region of the reference, so that the mapper can examine them as a single candidate. \textit{Chaining} links seed hits that appear in the same order and at compatible distances in both the read and the graph, and scores each chain with DP, so that only the highest-scoring chains are extended~\cite{ma2023chaining,chandra2023sequence}.\\
\noindent\textbf{Extension.} This final step expands the non-discarded seed hits into a candidate \textit{seed hit subgraph}, gathering enough neighboring nodes to cover the read, and then performs base-level alignment between the read and the subgraph's paths to produce the final mapping~\cite{smith1981identification,Cui2025}. The alignment proceeds in two phases. \nmg{First,} \textit{distance calculation} (DC) computes the edit distance between the read and a candidate reference path, i.e., the smallest number of edits that can explain their differences. \nmg{Second,} \textit{traceback} (TB) then recovers the \textit{actual alignment}, i.e., the sequence and position of the individual matches, substitutions, insertions, and deletions of the read with respect to the candidate path that achieve this minimum number of edits~\cite{cali2020genasm,li2009sequence}. If the edit distance computed by DC exceeds a predefined \textit{edit-distance threshold}, the candidate cannot yield a valid alignment and is rejected before TB is performed. In S2G mapping, this process determines both the actual alignment and the path within the seed hit subgraph that yields it, i.e., the \textit{valid alignment path}. We call each alignment of the read against one path of a seed hit subgraph a \textit{candidate alignment}. A candidate whose edit distance does not exceed the threshold is a \textit{valid alignment}; otherwise, it is \textit{discarded}. The best valid alignment of a read is its \textit{final alignment}, and the subgraph that contains it is the \textit{final alignment subgraph}.
Because graph alignment is computationally expensive, several optimized algorithms and hardware designs have been proposed to accelerate the end-to-end mapping workflow or individual steps (e.g.,~\cite{rautiainen2020graphaligner,kim2019hisat2,gao2020abpoa,jain2019pasgal,Rautiainen2019,Chandra2023,Ivanov2022,ma2023chaining,Darby2020vargas,Hwang2025MEMO,Romain2023svjedi,Li2020minigraph,Cali2022SeGraM,Zhang2024,Zeng2024asgdp,Shen2024128parallel,Li2024,Mandal2020,Varma2013,Awan2021,Feng2021,Zhang2025,kim2025nmp,Huang2023meg2,Angizi2020Panda,Qiu2017}).

\section{Motivational Analysis}
\label{sec:motivation}

We profile state-of-the-art S2G mapping tools to identify their main inefficiencies. We show that \inum{i} most of their costly alignment operations are performed on candidate alignments that do not yield valid alignments, and \inum{ii} these redundant operations dominate end-to-end mapping time. We profile the state-of-the-art software mappers vg~\cite{garrison2018variation} and GraphAligner~\cite{rautiainen2020graphaligner} and the state-of-the-art S2G mapping accelerator SeGraM~\cite{Cali2022SeGraM} (see our detailed profiling and evaluation methodology in \secref{sec:eval-meth}).

\subsection{Observation: Redundant Alignment Work}

\begin{figure}[h]
    \centering
    \includegraphics[width=0.9\columnwidth]{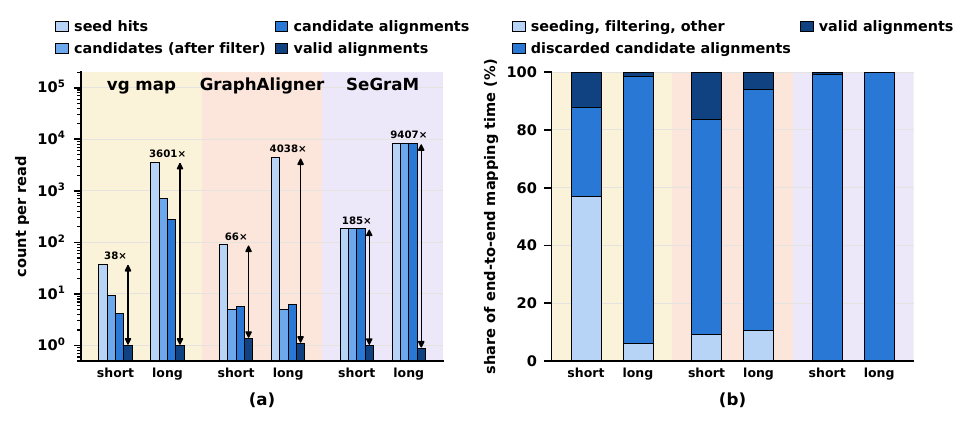}
\caption{(a)~Seed hits vs.\ valid alignments per read and (b)~end-to-end mapping time breakdown of S2G tools.}
    \label{fig:seed2align}
\end{figure}

\fig{\ref{fig:seed2align}(a)} shows the number of seed hits, candidate alignments, candidate alignments after filtering, and valid alignments per read for short reads and long reads. For short reads, vg, GraphAligner, and SeGraM find 38$\times$, 66$\times$, and 185$\times$ more seed hits than valid alignments, respectively. For long reads, this gap grows to 3601$\times$, 4038$\times$, and 9407$\times$, respectively. Long reads widen the gap for two reasons. First, a long read contains many more seeds and thus produces many more seed hits, yet it still has only one or a few valid alignments. Second, a long read spans many more variant sites, so its seed hit subgraphs contain many more alternative paths. The chaining and alignment steps need to examine these paths, further contributing to potentially redundant costly operations.

Each seed hit that survives filtering (i.e., candidates after filtering in \fig{\ref{fig:seed2align}}) triggers costly DP-based operations, such as chaining and alignment. As a result, nearly all of this work is spent on candidates that are eventually discarded. Notably, vg and GraphAligner already filter, cluster, and chain seed hits, yet they still examine orders of magnitude more candidates than needed. Thus, existing filtering heuristics cannot eliminate redundant alignments.

\textbf{Impact on End-to-End Mapping Time.} Alignment dominates the runtime of all three tools, as it takes 43-96\% and 89-91\% of the mapping time of vg and GraphAligner, respectively, and all of SeGraM's mapping time.\footnote{SeGraM pipelines seeding with its alignment unit, fully hiding seeding latency.} As \fig{\ref{fig:seed2align}(b)} shows, operations on discarded candidates alone account for 31-99.3\% of the end-to-end mapping time for short reads and 83-99.9\% for long reads across the tools. Therefore, identifying which candidates can lead to a valid alignment \emph{before} aligning them can drastically reduce the end-to-end mapping time.

\subsection{Opportunity: Recurring Topologies}
\label{sec:motifs}

\begin{figure}[h]
    \centering
    \includegraphics[width=0.9\columnwidth]{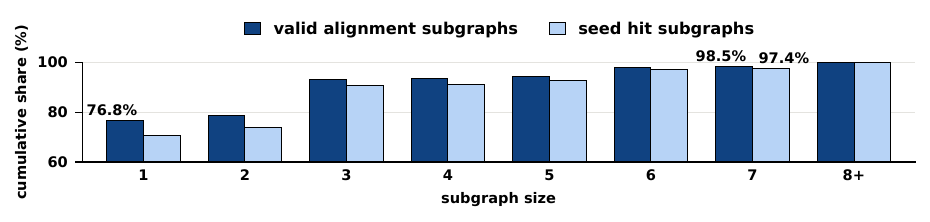}
    \caption{Cumulative percentages of the sizes of seed hit and valid alignment subgraphs.}
    \vspace{-1mm}
    \label{fig:subgraph_size_dist}
\end{figure}

Most genetic variation in genome graphs consists of SNPs and small indels~\cite{Liao2023}, which form \emph{small hops}, i.e., short branches that rejoin the shared backbone within a few nodes~\cite{garrison2018variation,paten2017genome}. As a result, many subgraphs share the same topology. This is an opportunity: if a few topologies describe most of the graph, the paths around a seed hit can be determined without traversing the graph to find all candidate alignment paths at runtime. 
\fig{\ref{fig:subgraph_size_dist}} shows that, for short reads, 98.5\% of valid alignment subgraphs and 97.4\% of seed hit subgraphs in SeGraM have fewer than 8 nodes. For long reads, 96.8\% of seed hit subgraphs decompose into chains of such small subgraphs separated by only 1-2 nodes. Thus, subgraphs of up to 7 nodes cover almost every subgraph that S2G mapping processes, either entirely or in parts.

\begin{figure}[h]
    \centering
    \includegraphics[width=0.9\columnwidth]{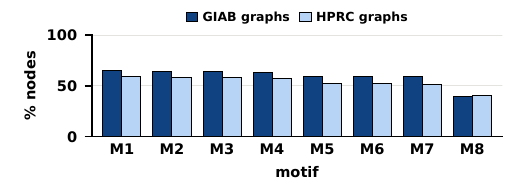}
    \caption{\revcqtwo{Occurrence of the most frequent motifs (\% of nodes at which each motif starts) in the GIAB and HPRC graphs. %
    }}
    \label{fig:motif_distribution}
\end{figure}

We call these small recurring topologies \emph{motifs}. A motif spans $n$ consecutive nodes, starting at the first and ending at the last node, and specifies which of the nodes in between it visits. We measure the occurrence of each motif ($3\leq n\leq7$) as the percentage of graph nodes at which it starts. \fig{\ref{fig:motif_distribution}} shows that motifs are highly recurrent. 
Therefore, a small set of motifs suffices to describe the topology of genome graphs of different densities.

\noindent\textbf{Our goal} is to improve the performance and energy efficiency of end-to-end S2G mapping by eliminating redundant alignment operations. To this end, we leverage motifs to identify, before alignment, the few paths around each seed hit that lead to a valid alignment.

\providecolor{darkgreen}{rgb}{0.0, 0.39, 0.0}
\providecommand\dg[1]{{\color{darkgreen}{#1}}}
\providecommand\tgmasic{\textsf{\toolname{}-ASIC}\xspace}
\providecommand\tgmcpu{\textsf{\toolname{}-CPU}\xspace}
\section{\tgm{}: High-Level Overview}
\label{sec:overview}

\begin{figure}[h]
    \centering
    \includegraphics[width=0.8\columnwidth]{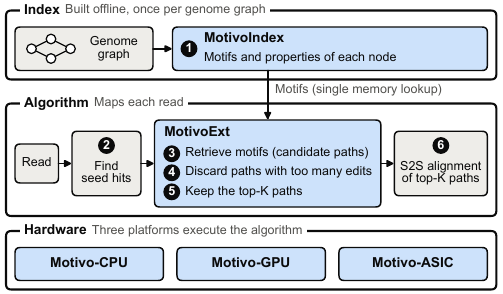}
    \caption{\textcolor{black}{High-level overview of \tgm{}.}}
    \label{fig:overview_konkanello}
\end{figure}

\klgh{We propose \tgm{},} \dg{a} \klgh{topology-aware algorithm-architecture co-design for sequence-to-graph (S2G) mapping}\dg{.
The key idea of \tgm{} is to use \emph{motifs} to identify, \emph{before} alignment, the few paths around a seed hit that can lead to a valid alignment.
A motif is a path over $n$ consecutive nodes of the genome graph ($3\leq n\leq7$, \sect{\ref{sec:motifs}}).
\fig{\ref{fig:overview_konkanello}} shows the three key components of \tgm{} (an index, an algorithm, and three hardware implementations of the algorithm) and the six steps in which \tgm{} works (\circled{1}-\circled{6}).}

\dg{\textbf{Building the index.} \tgm{} builds \tgmindex{} from the genome graph once, offline (\circled{1}, \sect{\ref{sec:ref-graph-index}}). \tgmindex{} is a compact graph index that stores, for each node of the genome graph, the precomputed motifs that start or end at the node and the properties of the node (i.e., its sequence length and its base composition, which is the number of A, C, G, and T bases in its sequence).}

\dg{\textbf{Mapping a read.} The algorithm first finds the seed hits of the read (\circled{2}). For each seed hit, \tgmext{} (\sect{\ref{sec:workflow}}) then selects the few most suitable mapping paths in three steps.}
\dg{First,} \tgmext{} retrieves \dg{the motifs that start (end) at the node of the seed hit for a right (left) extension} from \tgmindex{} \dg{with a single memory lookup (\circled{3}). Each motif is one candidate path around the seed hit, so \tgmext{} obtains the candidate paths \emph{without} traversing the graph.}
\dg{Second, \tgmext{} discards the candidate paths that cannot lead to a valid alignment (\circled{4}). \tgmext{} uses the sequence lengths of the nodes to find the part of each path that the read covers,} \nmg{sums the} precomputed base \nmg{compositions of the nodes} of \dg{this part,} \nmg{and compares the result with} \dg{the base composition of the read. The sum of the differences between the four base counts, i.e., the \emph{ACGT distance}, gives a lower bound on the number of edits of the path, because one edit changes the ACGT distance by at most two. \tgmext{} discards a path whose lower bound exceeds the edit-distance threshold, so this step discards no path that can lead to a valid alignment.}
\dg{Third, \tgmext{} keeps the top-$K$ paths, i.e., the $K$ paths with the smallest lower bounds (\circled{5}; we use $K=2$). This step is a heuristic; \sect{\ref{sec:accuracy}} quantifies its effect on the mapping accuracy.}
\dg{Finally,} \nmg{\tgm{} constructs the sequence}s of \dg{the top-$K$} paths \nmg{only} \dg{and aligns the read against each of them (\circled{6}). Thus, \tgmext{}} \nmg{reduces S2G mapping to a few} \dg{sequence-to-sequence (S2S)} \nmg{alignments per seed hit}\dg{, and dynamic programming (DP) remains only in the S2S aligner.}

\dg{\textbf{Executing the algorithm.} \tgm{}'s algorithm enables \inum{i}~high-performance execution on two conventional hardware platforms (\tgmcpu{} and \tgmgpu{}) and \inum{ii}~high-performance and low-energy execution on a new algorithm-architecture co-designed ASIC-based accelerator (\tgmasic{}, \sect{\ref{sec:arch}}).
\tgmcpu{} and \tgmgpu{} are straightforward software implementations of the algorithm on a CPU and on a GPU. They outperform highly optimized state-of-the-art CPU-based and GPU-based tools (\sect{\ref{sec:cpu-gpu}}), which shows that the benefits of \tgm{} come from the algorithm itself.
\tgmasic{} executes the algorithm end to end in hardware. The algorithm enables a lightweight design of \tgmasic{} in two ways.
First, since the path selection of the algorithm (steps \circled{3}-\circled{5}) uses only memory lookups and arithmetic operations, \tgmasic{} consists of simple hardware units, organized as a fully pipelined, five-stage dataflow.
Second, \tgmindex{} encodes each motif as a bitmask, and the set of all motifs fits in a small on-chip memory.
Thus, \tgmasic{} provides \textcolor{black}{high throughput per mm$^2$ and per Watt} (\sect{\ref{sec:s2g-accel}} and \sect{\ref{sec:energy}}).
}

\section{\tgm{}: Detailed Design}

\subsection{Data Structures}
\label{sec:ref-graph-index}
In this section, we present our novel motif-based indexing scheme, \tgmindex.
\klgh{The end goal of \tgmindex is to enable fast retrieval of precomputed paths around the seed hits to avoid expensive graph traversal at mapping time. To this end, \tgmindex encodes, for each node, the motifs it participates in, so that all candidate paths around the node are retrieved with a single lookup.}
\tgmindex{} \klgh{consists of two parts: \inum{i} a motif-based index that stores the motifs each node initiates (terminates), and \inum{ii} a compact structure that stores \textit{node properties} (e.g., sequence length, base composition)}. 
\klgh{\tgmindex fully exploits the motif recurrence by grouping nodes that initiate (or terminate) the same set of motifs. This enables compact, resource-efficient, and thus hardware-friendly storage of \emph{only} an identifier (ID) of the group the nodes belong to.}

\textit{Before indexing:} We topologically sort the graph using Kahn's algorithm, 
\klgh{similarly to state-of-the-art S2G accelerators and aligners~\cite{Cali2022SeGraM,Zhang2024,pasgal_data}.} \labelf{3} \klgh{\tgm{} supports graphs of various sizes and densities (e.g., GIAB and HPRC graphs), and does not require the graph to be acyclic.} \klgh{For pangenome graphs that contain cycles, such as HPRC, we use the node order of the sorted graph produced by vg, and store the few backward edges as special edges}. Alongside \tgmindex, we also build a \textit{minimizer-based index} ($k=15, w=20$), 
as described in~\sect{\ref{sec:background}}, needed during the seeding step. %

\begin{figure}[h]
    \centering
    \includegraphics[width=0.9\columnwidth]{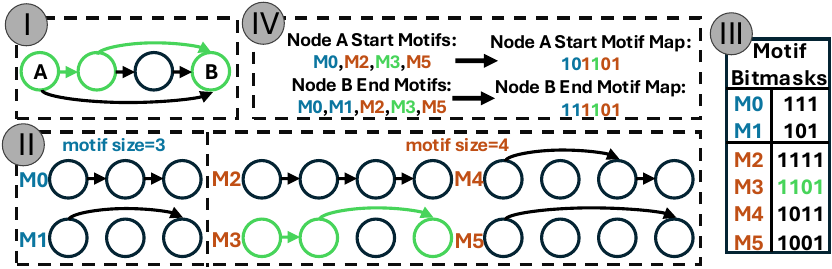}
    \caption{\tgmindex motif-based index example.}
    \label{fig:index_example}
\end{figure}

\textbf{Motif-based Index.} 
\klgh{\tgmindex}  \inum{i} encodes motif topologies as \textit{motif bitmasks}, \inum{ii} records the motifs each node initiates (terminates) in \textit{motif maps}, and \inum{iii} groups nodes with identical motif maps,\klgh{ to retrieve a node's motifs with a single lookup.}
\klgh{We describe these structures and use as an example a four-node graph (\fig{\ref{fig:index_example}}).}\\ 
\textit{Motif Bitmasks.} A motif of size $n$ is a path from the start to the end node of an $n$-node subgraph. Based on our profiling (\sect{\ref{sec:motivation}}), we enumerate all motifs with $3\leq n\leq7$. Each motif is encoded as an $n$-bit \textit{motif bitmask} (\fig{\ref{fig:index_example}}), where bit $i$ is set if the $i$-th node is on the path. Bits 0 and $n-1$ (start and end nodes) are always set. In total, there are $N=\sum_{i=3}^{n} 2^{i-2} = 2^{n-1}-2$ motifs. 
\klgh{In \fig{\ref{fig:index_example}}, for the graph of size $n=4$ (I) with start node $A$ and end node $B$, we depict all motifs $M_0$--$M_5$ it may include (II). The table (III) shows their bitwise encoding into $4$-bit bitmasks.} 
\klgh{For $n=7$, the number of motifs is 62.} \\
\textit{Motif Maps.} %
\klgh{Each node has a start and an end \textit{motif map}, which encode the motifs that the node initiates (\textit{start motifs}) and terminates (\textit{end motifs}), respectively. The length of the \textit{motif map} is equal to the number of motifs, i.e., 6 in \fig{\ref{fig:index_example}} and 62 in the full \tgmindex{}. The $j$-th bit of the motif map is set if the node initiates (terminates) motif $M_j$. \fig{\ref{fig:index_example}} shows the start and end motif maps (IV) for nodes A and B, respectively.} %

\begin{figure}[h]
    \centering
    \includegraphics[width=0.9\columnwidth]{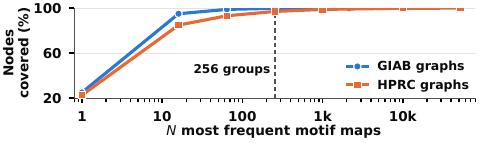}
    \caption{Motif Groups Node Coverage.}
    \label{fig:map_coverage}
\end{figure}

\textit{Node Motif-based Grouping.} After creating start and end motif maps for all nodes in the graph, we leverage the recurrence of motifs (observed in~\sect{\ref{sec:motifs}}) to reduce memory capacity requirements. More specifically, as observed in \fig{\ref{fig:map_coverage}}, \revcqtwo{a few motif maps characterize most nodes in both graph types: the most frequent start motif map alone covers 23-25\% of all nodes, while the 256 most frequent ones cover 99.85\% (98.1\%) of the nodes of the GIAB (HPRC) graphs.} This enables us to group the nodes based on motif maps and store for each node the group ID of the group it belongs to. We also do the same grouping based on end motif maps.

\textbf{Node Properties.} This part of \tgmindex stores \textit{node properties} needed by \tgmext and the alignment step. For each node $i$ of the graph, \klgh{we store} its actual sequence, its sequence length, and its ACGT-content (i.e., the count of each nucleotide base in the node sequence).

\textbf{Special cases.}\labelcqtwo{2} \klgh{In the case of complex, higher-degree graphs that include structural variations and cycles, some nodes do not belong to any of the frequent groups. To account for those, \emph{without} accessing the graph's adjacency lists during mapping, we introduce two bits for each node}: \inum{i} the \textit{rare topology} bit and \inum{ii} the \textit{special edge} bit. If the \textit{rare topology} bit of a node is set to one, that node is characterized by a rare combination of motifs (i.e., its group is not among the most common ones), and its motif map is kept instead of a group ID. Only 0.0027\% (0.0138\%) of the nodes of our GIAB (HPRC) human genome graphs have rare topologies. 
Since a motif can only represent edges between nearby nodes in the sorted order, we introduce the \textit{special edge bit} to compactly store the remaining few edges, i.e., edges to far-away nodes (e.g., due to structural variants) and backward edges. We use a list of far or backward neighbors of the nodes flagged with the special edge bit. Only 0.18\% (3.17\%) of the nodes of our GIAB (HPRC) human genome graphs have a far-away neighbor, while only 1.23\% of the HPRC graphs' nodes have a backward neighbor.
\reve{\labele{8}Across all 24 chromosomes, the size of the complete index (\tgmindex{} and minimizer index) is 8.15\,GB (14.25\,GB) for the GIAB (HPRC) graphs. }

\subsection{S2G Mapping Algorithm}
\label{sec:workflow}
\begin{figure}[h]
    \centering
    \includegraphics[width=0.9\columnwidth]{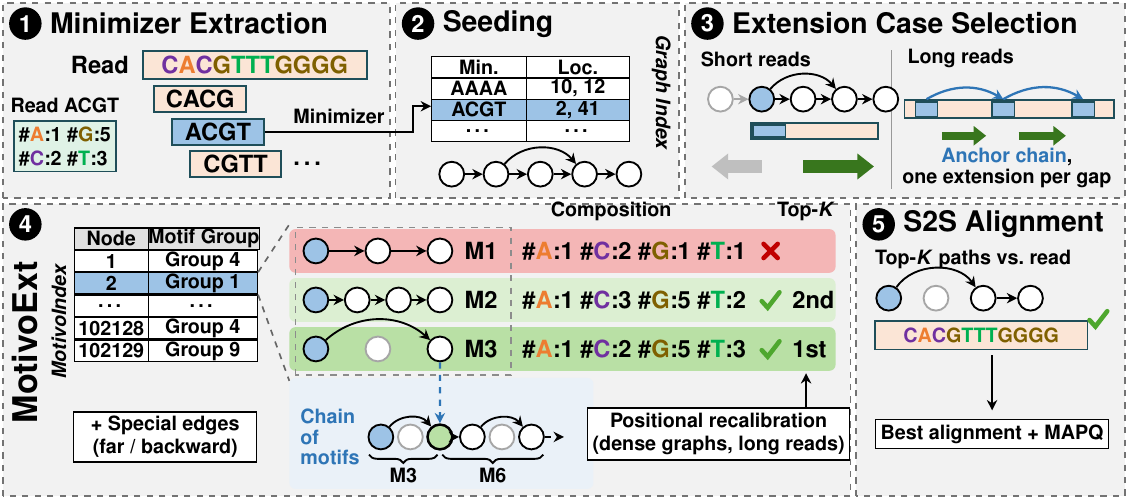}
    \caption{\tgm{} end-to-end S2G mapping workflow.}
    \label{fig:tagma_overview_workflow}
\end{figure}

\noindent\textbf{Step-by-step flow.}
\fig{\ref{fig:tagma_overview_workflow}} shows \tgm{}'s end-to-end S2G online read mapping algorithm, which consists of five steps: \inum{1} minimizer extraction, \inum{2} seeding, \inum{3} extension case selection, \inum{4} extension, and \inum{5} S2S alignment. First, \textit{Minimizer Extraction}~\tikz[baseline=(char.base)]{\node[shape=circle, draw, fill=black, text=white, inner sep=1.5pt] (char) {1};} extracts minimizer subsequences from the read sequence. Second, \textit{Seeding}~\tikz[baseline=(char.base)]{\node[shape=circle, draw, fill=black, text=white, inner sep=1.5pt] (char) {2};} uses these minimizers to query the minimizer-based index and find seed hit locations in the graph. Third, \textit{Extension Case Selection}~\tikz[baseline=(char.base)]{\node[shape=circle, draw, fill=black, text=white, inner sep=1.5pt] (char) {3};} determines which extension direction is needed to have adequate nucleotides to align against the read. Then, \textit{\tgmext}~\tikz[baseline=(char.base)]{\node[shape=circle, draw, fill=black, text=white, inner sep=1.5pt] (char) {4};} queries \tgmindex{} to retrieve the motifs the seed hit node initiates or terminates, depending on the selected extension. \tgmext{} uses the motif topologies and the nodes' base composition to determine which motif represents the \revcqone{paths most likely to yield valid alignments}. Finally, \textit{S2S Alignment}~\tikz[baseline=(char.base)]{\node[shape=circle, draw, fill=black, text=white, inner sep=1.5pt] (char) {5};} \revcqone{aligns these paths' sequences against the read and produces the final alignment and its mapping quality (MAPQ).}
\noindent\textbf{Minimizer Extraction.} The first step extracts the read's minimizers, using the same hash function and parameters ($k$, $w$) 
as the minimizer-based index, and counts its bases to obtain the read's ACGT-content for later stages. \revcqfour{Long reads are divided into non-overlapping windows, for which we compute the ACGT-content and sample minimizers with density $D$ to bound the seeding work.}
\noindent\textbf{Seeding.} The seeding step queries the read's minimizers against the minimizer-based index.
Each seed hit keeps its graph position (node ID and offset in the node sequence) and its offset in the read.
\noindent\textbf{Extension Case Selection.}
\label{sec:alg_case}
This step leverages the positional information of seed hits to determine whether and in which direction(s) the seed hit subgraph must be extended.  \klgh{The goal is for the subgraph to \emph{span enough characters to align the read in full}, based on the offsets of the seed hit in the reference and read sequences.} There are four cases of possible extension direction: \inum{i} No extension~\klgh{($NE$)}, i.e., the seed hit subgraph includes only the seed hit node, \inum{ii} Left extension~\klgh{($LE$)}, i.e., the seed hit subgraph needs to include nodes to the left of the seed hit node, \inum{iii} Right extension~\klgh{($RE$)}, i.e., the seed hit subgraph needs to include nodes to the right of the seed hit node, and \inum{iv} Bidirectional extension~\klgh{($BE$)}, i.e., both left and right extension are applied. \klgh{For long reads, this step performs \textit{chaining} before extension. Chaining links seed hits that are close to each other in both the read and the graph into a chain of anchors.}
\noindent\textbf{Extension.}
\label{sec:alg_ext}
The extension step extends seed hits to seed hit subgraphs and filters out subgraphs that cannot lead to valid alignments. It handles case $NE$ with a basecounting (BC) filter and cases $LE, RE, BE$ with \tgmext. All cases use \tgmindex{} and the ACGT distance, $\sum_{i\in\{A,C,G,T\}} \left| SeqA\_content_i - SeqB\_content_i \right|$ between the read and the candidate alignment sequences. The minimum number of edits between two sequences is $\lceil ACGT\_distance/2 \rceil$. When no extension is needed (i.e., the reference subsequence lies entirely within the seed hit node), we fetch the subsequence from \tgmindex, and the BC filter discards it if its minimum number of edits to the read exceeds the edit-distance threshold.
For multi-node seed hit subgraphs, \tgmext first retrieves the start (end) motifs of the seed hit node from \tgmindex for right (left) extension. Second, it uses the nodes' sequence lengths to find the exact end (start) offset in each motif's end (start) node. Third, it adds the ACGT-content of that node's segment to the precomputed ACGT-content of the motif's intermediate nodes, and discards motifs whose minimum number of edits exceeds the edit-distance threshold. \revcqtwo{Fourth, for nodes long enough to lie fully within the read, \tgmext performs a lightweight positional recalibration, i.e., a short exact-match comparison between the node's sequence and the read around the expected position. This keeps the read on track with the path, so the ACGT-content is compared only over the short read segments between such positions, where the candidate paths actually differ. For long reads, where many alternative paths of a variation site often have identical ACGT-content, \tgmext also compares both of them directly against the read with bitwise operations. These features keep \tgmext robust to denser graphs and noisier reads using only comparisons and additions.}
\revcqone{Finally, \tgmext keeps the top-$K$ candidate paths, ranked by their minimum number of edits;} \reve{candidates that are tied or within one edit of each other are re-ranked by their exact edit distance to a short read segment, using part of the S2S aligner.\labele{6}} These paths form the top-$K$ seed hit subgraphs, i.e., the \textit{valid alignment motifs}. \revcqone{If no motif of the seed hit node covers the needed number of bases, \tgmext continues from the last node of each kept path, building the path as a chain of motifs. For rare nodes within a motif, it continues from their far or backward neighbors.} \revcqfour{For long reads, the same procedure extends each anchor until it reaches the next anchor of the chain.}

\noindent\textbf{Alignment.}
\label{sec:alg_aligner}
Each candidate path is likely to produce a valid alignment with the read sequence. We fetch the sequences of its nodes, concatenate them into a single linear sequence, and align it against the read with the bitwise operation-based alignment algorithm GenASM~\cite{cali2020genasm}. The best alignment over all candidate paths gives the mapping, i.e., the path coordinates and the CIGAR string, i.e., a compact encoding of the matches, mismatches, insertions, and deletions of the alignment. \reve{Since \tgm{} keeps the top-$K$ paths, it also keeps the best alignments of the read's top loci and computes the MAPQ from the difference between the best and second-best alignments, as in conventional mappers~\cite{li2018minimap2,Jung_2022}.}

\subsection{Hardware for S2G Mapping}
\label{sec:arch}
\begin{figure}[h]
    \centering
    \includegraphics[width=0.9\columnwidth]{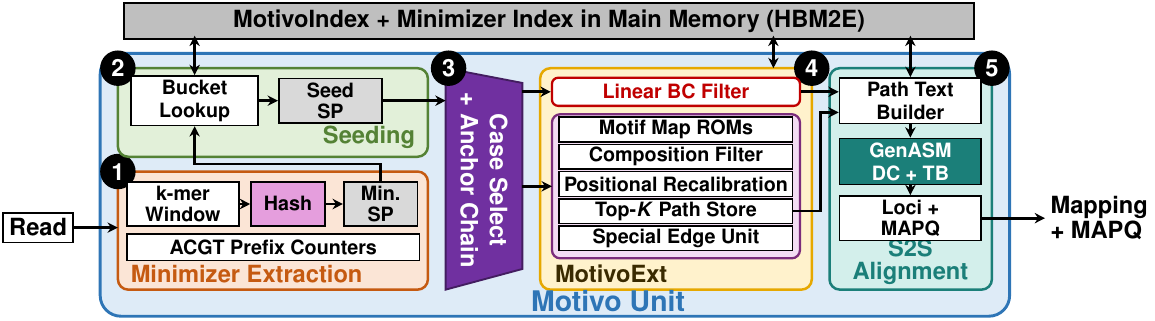}
    \caption{\tgm Hardware for S2G Mapping.}
    \label{fig:tagma_overview}
\end{figure}

\tgm{}'s hardware design for S2G mapping is implemented as a fully pipelined, five-stage dataflow with inter-stage double buffering.
\klgh{\fig{\ref{fig:tagma_overview}} shows a high-level overview of the design, which comprises five modules:} 
the \textit{Minimizer Extraction Module}~\tikz[baseline=(char.base)]{\node[shape=circle, draw, fill=black, text=white, inner sep=1.5pt] (char) {1};}, the \textit{Seeding Module}~\tikz[baseline=(char.base)]{\node[shape=circle, draw, fill=black, text=white, inner sep=1.5pt] (char) {2};}, the \textit{Extension Case Selection Module}~\tikz[baseline=(char.base)]{\node[shape=circle, draw, fill=black, text=white, inner sep=1.5pt] (char) {3};}, the \textit{\tgmext{} Module}~\tikz[baseline=(char.base)]{\node[shape=circle, draw, fill=black, text=white, inner sep=1.5pt] (char) {4};}, and the \textit{S2S Alignment and Loci Module}~\tikz[baseline=(char.base)]{\node[shape=circle, draw, fill=black, text=white, inner sep=1.5pt] (char) {5};}.

\subsubsection{Minimizer Extraction}
\label{sec:arch_minimizer}
Minimizers are extracted in a single pass by a rolling window, using a $w$-entry circular buffer that holds the current window's k-mers and a combinational minimizer finder that selects one minimizer per cycle. Each minimizer and its read offset are written to an on-chip scratchpad (Min. SP), so the next minimizer is prefetched while the Seeding block queries the previous one. The positional offset is forwarded to module~{\tikz[baseline=(char.base)]\node[shape=circle, draw, fill=black, text=white, inner sep=1.5pt] (char) {3};} if a seed hit is found.
\subsubsection{Seeding}
\label{sec:arch_seeding}
The minimizer-based index is a hash table of fixed-size buckets, so each lookup takes a single main-memory access. Seed hit positions are written to an on-chip scratchpad (Seed SP).
\subsubsection{Extension Case Selection}
\label{sec:arch_case}
Module~{\tikz[baseline=(char.base)]\node[shape=circle, draw, fill=black, text=white, inner sep=1.5pt] (char) {3};} is a simple combinational circuit that compares the seed hit's offsets within the read and within the \klgh{seed hit} node against the node's sequence length, and decodes the outcome into the $enable$ signals of the corresponding extension modules. \klgh{The chaining for long reads is implemented through a small controller that builds the anchor chain using simple comparisons of node IDs and positional distances.}

\subsubsection{Extension}
\label{sec:arch_ext}
\begin{figure}[h]
    \centering
    \includegraphics[width=0.9\columnwidth]{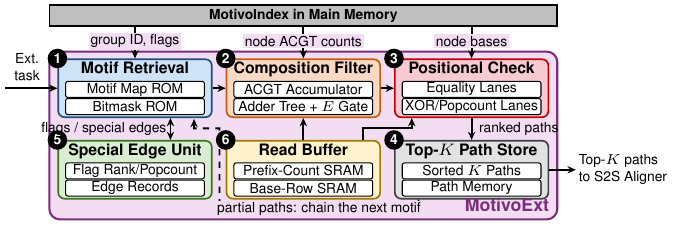}
    \caption{\tgmext{} Micro-Architecture.}
    \label{fig:tagma_ext}
\end{figure}

This module includes the \tgmext block and the simple linear basecounting filter, which is applied in the no-extension case. The basecounting filter performs simple base counting, arithmetic operations to calculate ACGT distance, and comparisons to discard a read/reference sequence pair based on the provided edit-distance threshold. \\
\textbf{\textit{\tgmext Module.}} At the core of the extension block lies the \tgmext module. \fig{\ref{fig:tagma_ext}} shows the hardware architecture of the \tgmext block, which performs the motif-based extension of the seed hits.
\textit{Motif Retrieval}~\tikz[baseline=(char.base)]{\node[shape=circle, draw, fill=black, text=white, inner sep=1.5pt] (char) {1};}. Thanks to the intrinsic properties of the genome variation graphs, including the existence of known motifs and their repeatability, we can efficiently keep their bitmasks and the group motif maps in on-chip ROMs to reduce off-chip memory accesses. To minimize the area of that ROM, we leverage the following observation: \revcqtwo{the 256 most common groups cover 99.85\% (98.1\%) of the nodes of our GIAB (HPRC) graphs}; thus we only need a 4\,KB on-chip memory to store them (for start and end groups), \revcqtwo{selected by the extension direction}. Since we only have 62 7-bit motif bitmasks, we store them in a small ROM. \revcqtwo{Each node's group ID, rare and jump bits, sequence length, and ACGT-content are packed into a 16-byte record, so a single memory access retrieves both its topology and its properties. Since the records are stored in topological order, the nodes of a motif fall in one or two adjacent 64-byte lines, which a small line cache exploits. Only for rare nodes is the node's own motif map fetched from main memory.} Representing motif maps and motif bitmasks in a bitwise manner enables detecting the present motifs and their nodes through simple bit shifting and isolation of the LSB.
\textit{Composition Filter}~\tikz[baseline=(char.base)]{\node[shape=circle, draw, fill=black, text=white, inner sep=1.5pt] (char) {2};}. The ACGT accumulator adds the precomputed ACGT-content of the motif's nodes and, since only part of the sequence of a boundary node may be needed, counts the needed part with masked popcounts over the 2-bit encoded sequence. The \textit{Read Buffer}~\tikz[baseline=(char.base)]{\node[shape=circle, draw, fill=black, text=white, inner sep=1.5pt] (char) {6};} stores the read's ACGT prefix counts, so that the ACGT-content of any read segment is obtained with a single subtraction. An adder tree computes the minimum number of edits, and a comparator discards the motifs that exceed the edit threshold.
\revcqone{\textit{Positional Recalibration}~\tikz[baseline=(char.base)]{\node[shape=circle, draw, fill=black, text=white, inner sep=1.5pt] (char) {3};} uses parallel equality comparators to locate long nodes in the read and re-anchor the read. For long reads, XOR and popcount lanes compare alternative alleles against the read. The \textit{Top-K Path Store}~\tikz[baseline=(char.base)]{\node[shape=circle, draw, fill=black, text=white, inner sep=1.5pt] (char) {4};} keeps the top-$K$ paths in a small sorted buffer and feeds the partial ones back to block~\tikz[baseline=(char.base)]{\node[shape=circle, draw, fill=black, text=white, inner sep=1.5pt] (char) {1};} to chain their next motif.} \revcqtwo{The \textit{Jump Unit}~\tikz[baseline=(char.base)]{\node[shape=circle, draw, fill=black, text=white, inner sep=1.5pt] (char) {5};} retrieves the far and backward neighbors of the nodes whose \emph{jump} bit (\sect{\ref{sec:ref-graph-index}}) is set.} \revcqone{Because only the top-$K$ paths are passed to the S2S alignment step, the aligner is invoked a few times per extension rather than once per candidate, concentrating the expensive alignment on the most promising subgraphs.}

\subsubsection{Alignment}
\label{sec:arch_aligner}
\textcolor{black}{In order to build the sequence of the top-$K$ paths for the alignment step, the Path Text Builder module fetches and concatenates the node sequences of each candidate path from memory and passes the result} to the S2S alignment block, which is based on GenASM-DC/TB~\cite{cali2020genasm}. GenASM-DC executes the bitvector DP using bitwise operations (e.g., AND/OR/XOR/shift), and GenASM-TB traverses the stored DP bitvectors to recover the alignment. \reve{A small scratchpad keeps the best alignment of each of the top loci, and a lookup table converts their edit-distance difference into the MAPQ score.}

\section{Evaluation Methodology}
\label{sec:eval-meth}

\revcqtwo{\textbf{Datasets.} We evaluate two groups of graphs. \inum{i} \textbf{GIAB graphs}: as in prior work~\cite{Cali2022SeGraM}, we build 24 graphs (chromosomes 1-22, X, and Y) from the GRCh38 reference~\cite{schneider2017evaluation} and the variants of HG001-HG007 from the GIAB project (v3.3.2)~\cite{zook2016extensive} with \texttt{vg construct}~\cite{garrison2018variation}. \inum{ii} \textbf{HPRC graphs}: to evaluate denser pangenome graphs, we use the per-chromosome Minigraph-Cactus graphs of the Human Pangenome Reference Consortium (HPRC)~\cite{Liao2023,Hickey2024MinigraphCactus}. 
For each chromosome and graph, we simulate \inum{i} three short-read sets with read lengths of 100, 150, and 250\,bp at a 1\% error rate (Illumina profile) with Mason~\cite{holtgrewe2010mason}, and \inum{ii} two long-read sets of 10\,kbp PacBio continuous long reads with PBSIM3~\cite{Ono2022PBSIM3}. We use the PacBio error model ERRHMM-RSII, with mean error rates of 5\% and 10\%. Each set contains 1M reads.}

\textbf{Evaluated Systems.}
We evaluate the \tgm{} algorithm on three platforms: \tgmcpu{} and \tgmgpu{} run it in software on a CPU and on a GPU, and \tgmasic{} runs it end to end in hardware. We compare each implementation against the state-of-the-art S2G mapping systems of its platform, summarized in Table~\ref{tab:systems}. We implement \tgmasic{} as an ASIC to ensure a fair comparison with the state-of-the-art ASIC-based S2G accelerators~\cite{Cali2022SeGraM,Zhang2024}.
\reva{\labela{2}We evaluate each system on all dataset types that it supports. We run the CPU-based systems on the same high-end server and the GPU-based ones on the same NVIDIA A100 GPU, each in its best-performing configuration.%
\footnote{Since vg-giraffe supports only short reads, we evaluate GPUs on short reads only.}} For SeGraM~\cite{Cali2022SeGraM} and \reva{\labela{1}HARP~\cite{Zhang2024}}, we use the area and power reported in their papers. SeGraM was evaluated only on the GIAB graphs, as its accuracy collapses for denser graphs, especially for long reads (<50\%), due to its low upper limit on subgraph size.
This setup separates the benefits of the \tgm{} algorithm from those of the hardware. First, since \tgmcpu{} and \tgmgpu{} run on the same platforms and datasets as their baselines, comparing \tgmcpu{} (\tgmgpu{}) with the CPU-based (GPU-based) baselines isolates the benefit of the algorithm. Second, since \tgmgpu{} produces the same mappings as \tgmcpu{}, and we validate \tgmasic{} against \tgmcpu{}, comparing the three implementations with each other isolates the benefit of the hardware.
For all systems, we measure the mapping step only, excluding offline graph construction and indexing, which take place once per reference and are not on the critical path of genome analysis~\cite{mansouri2025sage}. 
\begin{table}[t]
\centering
\caption{\protect\reva{Evaluated systems and datasets ($\surd$: measured by us; $\circ$: reported in the system's paper; --: not supported).}}
\label{tab:systems}
\footnotesize
\setlength{\tabcolsep}{2.5pt}
\begin{tabular}{@{}llcccc@{}}
\hline
 & & \multicolumn{2}{c}{\textbf{GIAB graphs}} & \multicolumn{2}{c}{\textbf{HPRC graphs}} \\
\cline{3-6}
\textbf{Platform} & \textbf{System} & \textbf{Short} & \textbf{Long} & \textbf{Short} & \textbf{Long} \\
\hline
CPU & GraphAligner~\cite{rautiainen2020graphaligner} & $\surd$ & $\surd$ & $\surd$ & $\surd$ \\
2$\times$ AMD EPYC 9554 & vg~\cite{garrison2018variation} & $\surd$ & $\surd$ & $\surd$ & $\surd$ \\
 & minigraph~\cite{Li2020minigraph} & -- & $\surd$ & -- & $\surd$ \\
 & \tgmsw & $\surd$ & $\surd$ & $\surd$ & $\surd$ \\
  & vg-giraffe & $\surd$ & -- & $\surd$ & -- \\
\hline
GPU & vg-giraffe~\cite{nvidia_parabricks_giraffe} & $\surd$ & -- & $\surd$ & -- \\
NVIDIA A100 80\,GB & \tgmgpu & $\surd$ & -- & $\surd$ & -- \\
\hline
ASIC & SeGraM~\cite{Cali2022SeGraM} & $\circ$ & $\circ$ & -- & -- \\
28\,nm & HARP~\cite{Zhang2024} & -- & -- & -- & $\circ$ \\
 & \tgmasic{} & $\surd$ & $\surd$ & $\surd$ & $\surd$ \\
\hline
\end{tabular}
\end{table}

\textbf{\tgm{} Performance.} We implement all of \tgm{} in synthesizable Verilog RTL, and validate the RTL against \tgmsw{} with Verilator~\cite{Snyder_Verilator}. We evaluate throughput with a cycle-level simulator fitted to the RTL cycle counts, \revd{\revmid{rd}{D4}which models the off-chip memory with Ramulator 2.1~\cite{ramulator_one,ramulator21source} (building on prior Ramulator versions~\cite{kim2016ramulator,ramulatorsource,luo2023ramulator,ramulator2source}).} \textcolor{black}{We use the simulator to enable fast evaluation of \tgm{} on large read datasets and various graphs.} \textcolor{black}{Our default configuration uses 32 \tgm{} units and a single HBM2E stack (8 channels, 16 pseudo-channels, 16\,GB). The reference index, whose size is up to 14.25\,GB for the larger graphs, is loaded once into the stack and shared by all units (see \sect{\ref{sec:scalability}} for more details). We choose HBM2E, the memory technology of SeGraM~\cite{Cali2022SeGraM}, so that the two accelerators are compared on the same memory technology.}

\noindent\textbf{Area, Power, and Energy.} We use Yosys~v0.33~\cite{wolf2013yosys} for \tgm{} synthesis (Nangate45 standard-cell library at a target clock
frequency of 1\,GHz), and perform physical design with the OpenROAD
open-source RTL-to-GDS flow~\cite{ajayi2019openroad} for post-layout
area and power.
For all \klgh{ASIC systems, we compare the area, power, and energy of the hardware compute units only}. \textcolor{black}{This comparison favors the baselines: HARP runs seeding on the host CPU, which is not included in its area and power, and SeGraM uses four HBM2E stacks with a replicated index, whereas \tgmasic{} uses a single stack.} 
\textcolor{black}{We model HBM power separately using Ayna~\cite{olgun2026understanding}, driven by the DRAM command traces of our Ramulator simulations and scaled from HBM2 to HBM2E.} For the CPU systems, we measure the energy counters of the processor packages. For the GPU systems, we measure the power of the GPU board with NVML~\cite{nvml}. 
For all systems, we measure performance, power, and energy consumed only during the mapping steps, excluding index loading.

\section{Evaluation Analysis}
\label{sec:evaluation-analyses}

\subsection{Performance}

\begin{figure}[h]
    \centering
    \includegraphics[width=0.9\columnwidth]{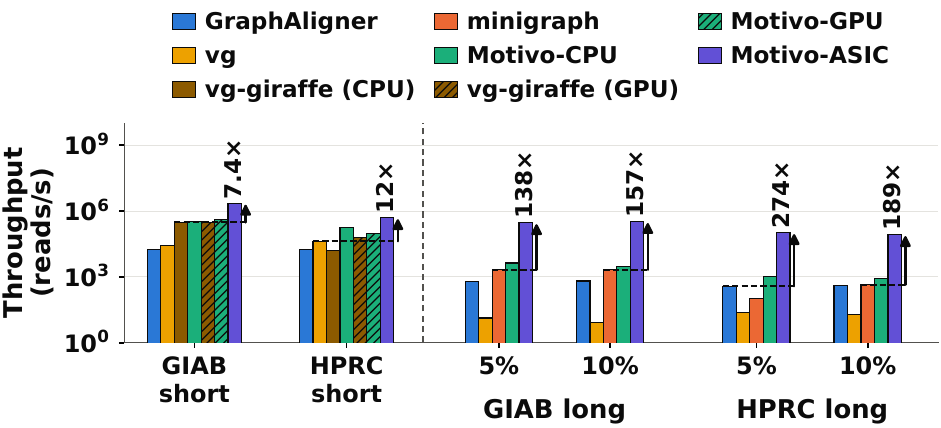}
    \caption{Throughput (reads/s) comparison with CPU and GPU baselines. }
    \label{fig:throughput_software}
\end{figure}

\fig{\ref{fig:throughput_software}} presents the throughput of the CPU- and GPU-based baselines and of the three implementations of \tgm{}: \tgmcpu{}, \tgmgpu{}, and \tgmasic{}. We analyze these results along two axes. First, we compare \tgmcpu{} and \tgmgpu{} with the state-of-the-art baselines on the same hardware, which shows the benefit of the \tgm{} algorithm (\sect{\ref{sec:cpu-gpu}}). Second, we compare the three implementations of \tgm{} with each other, which shows the benefit of the hardware (\sect{\ref{sec:platforms}}).
\label{sec:performance}
\subsubsection{Comparison with CPU- and GPU-Based Systems}
\label{sec:cpu-gpu}
We make two key observations. First, the algorithm of \tgm{} outperforms the state-of-the-art systems on the same hardware. On the CPU, \tgmcpu{} provides 2.1-18.4$\times$, 2.8-343$\times$, 1.1-7.1$\times$, and 1.4-10.1$\times$ speedups over GraphAligner, vg, vg-giraffe, and minigraph, respectively. On the GPU, \tgmgpu{} provides 1.5-1.6$\times$ speedups over vg-giraffe, with 1.7-1.9$\times$ lower energy per read (\fig{\ref{fig:energy}}). Second, the benefit of the algorithm grows with graph density. The denser HPRC graphs have more alternative paths around each seed hit. Thus, the baselines traverse the additional paths of denser graphs node by node, whereas \tgm{} fetches the precomputed paths through motif encodings.  For example, vg-giraffe performs close to \tgmcpu{} for short reads on the GIAB graphs, but \tgmcpu{} is 7.1$\times$ faster than vg-giraffe on the HPRC graphs.
\begin{figure}[h]
    \centering
    \includegraphics[width=0.9\columnwidth]{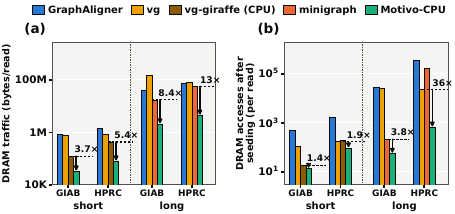}
    \caption{Per-read (a)~DRAM traffic and (b)~DRAM accesses after seeding, i.e., loads that miss all caches during candidate selection
    and alignment, of the CPU systems.}
    \label{fig:cpu_benefit}
\end{figure}

\noindent\textit{Sources of Improvement.} \fig{\ref{fig:cpu_benefit}} shows the per-read DRAM traffic and post-seeding DRAM accesses of the CPU-based systems. \tgmcpu{} has the lowest DRAM traffic across all read sets, 3.7-13$\times$ lower than the best-performing baseline (\fig{\ref{fig:cpu_benefit}a}), because \tgmindex{} replaces each graph traversal with a single memory lookup (\fig{\ref{fig:cpu_benefit}b}). This reduction grows with graph density. For long reads, \tgmcpu{} performs 3.8$\times$ fewer post-seeding DRAM accesses than minigraph on the GIAB graphs, and up to 36$\times$ fewer \textcolor{black}{than vg, the best-performing baseline} on the denser HPRC graphs. The reason is the increased number of graph traversals in the baselines. The baselines traverse the additional paths of denser graphs node by node, whereas \tgm{}'s motifs encode them compactly.

\subsubsection{Comparison of the \tgm{} Implementations}
\label{sec:platforms}
Since \tgmcpu{}, \tgmgpu{}, and \tgmasic{} run the same algorithm, comparing them with each other isolates the benefit of the hardware. We make two key observations.
First, general-purpose hardware limits the benefit of the algorithm. \tgmgpu{} is only 1.3$\times$ faster than \tgmcpu{} on the GIAB graphs, and slower on the HPRC graphs. After \tgmindex{} removes most memory accesses (\sect{\ref{sec:cpu-gpu}}), the remaining work is path selection. Path selection consists of lightweight lookups and arithmetic operations, but they form chains of short, dependent steps whose length varies widely across seed hits (\fig{\ref{fig:stage_breakdown}b}). \textcolor{black}{On the CPU, these dependent steps serialize on memory latency and branches, and on the GPU, the varying chain lengths leave only a few threads of a warp active.} General-purpose hardware thus spends its resources on control flow, memory latency, and divergence, rather than on the lightweight operations themselves.
Second, \tgmasic{} removes these limitations, which leads to large benefits. Since path selection uses only lookups and arithmetic operations, \tgmasic{} executes it with \tgmext{}, a simple dedicated unit that processes the dependent steps in a pipeline, without instruction overheads or divergence, and a small on-chip memory that holds all motifs as bitmasks, so that off-chip accesses are rare. With the same algorithm, \tgmasic{} is 4.2-6.9$\times$ (68-113$\times$) faster than \tgmcpu{} for short (long) reads, and 5.2-5.3$\times$ faster than \tgmgpu{}. The speedup is largest for long reads, whose extensions form the longest chains of dependent steps. As a result, \tgmasic{} provides 7.4-11.7$\times$ and 7.7-8.2$\times$ speedups over the state-of-the-art CPU- and GPU-based baselines for short reads, respectively, and 138-274$\times$ speedups over the state-of-the-art CPU-based baselines for long reads.
Ultimately, choosing among the three implementations is a design decision. \tgmcpu{} and \tgmgpu{} enable near-term adoption on existing hardware, whereas \tgmasic{} \inum{i}~provides the highest performance, \inum{ii}~achieves the highest energy efficiency (\sect{\ref{sec:energy}}), and \inum{iii}~\textcolor{black}{requires only 36.7\,mm$^2$ and 10.6-12.0\,W, i.e., less power than a single CPU socket or GPU of our baselines, which makes it a candidate for compact systems for on-site genome analysis}~\cite{MinIONMk1CO,palatnick2020igenomics,Ballard2018,Oehler2023,pomerantz2018real,chiang2019from,mutlu2023accelerating,alser2022molecules}.

\subsubsection{Comparison with S2G Accelerators}
\label{sec:s2g-accel}
For a fair comparison, we scale \tgmasic{}'s area and power down to the 28\,nm process technology node of SeGraM and HARP, following the methodology in~\cite{stillmaker2017Scaling}.
\begin{figure}[h]
    \centering
    \includegraphics[width=0.9\columnwidth]{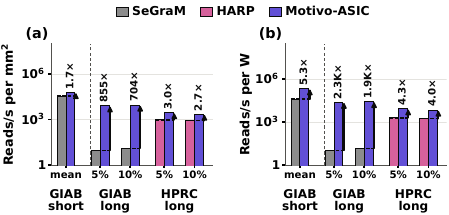}
    \caption{(a) Iso-area and (b) iso-power comparison of \tgmasic{} with SeGraM and HARP.}
    \label{fig:iso_area_power}
\end{figure}
\fig{\ref{fig:iso_area_power}} presents the throughput per mm$^2$ (iso-area) and per Watt (iso-power) of the three accelerators. We make two key observations.
First, on the GIAB graphs, \tgmasic{} outperforms SeGraM. \tgmasic{} provides \textcolor{black}{1.7$\times$} (5.3$\times$) higher throughput per mm$^2$ (Watt) for short reads and \textcolor{black}{704-855$\times$} (\textcolor{black}{1883-2286$\times$}) for long reads, and \textcolor{black}{1.9-2.7$\times$} (\textcolor{black}{932-1132$\times$}) higher raw throughput for short (long) reads.
Second, on the HPRC graphs, \tgmasic{} outperforms HARP. \tgmasic{} provides \textcolor{black}{2.7-3.0$\times$} higher throughput per mm$^2$ and \textcolor{black}{4.0-4.3$\times$} higher throughput per Watt than HARP for long reads. These gains come in addition to broader functionality: unlike HARP, \tgmasic{} also maps short reads and performs seeding on chip.

\noindent\textit{Sources of Improvement.} SeGraM aligns the subgraph of every seed hit, whereas \tgmasic{} selects the few most suitable paths with motifs before alignment. As a result, on the GIAB graphs, \tgmasic{} aligns 1.5-4.5$\times$ (243-521$\times$) fewer candidates, evaluates 1.2-9.7$\times$ (310-710$\times$) fewer DP cells, and issues 1.2-1.9$\times$ (36-85$\times$) fewer HBM accesses after seeding than SeGraM for short (long) reads. HARP aligns a similar number of candidates per long read (2.6-2.9 vs.\ 4.0-4.2), but each \tgmasic{} candidate is 2.0-3.0$\times$ cheaper, resulting in 1.4-2.0$\times$ fewer DP cells. Moreover, HARP requires a 1\,MB on-chip memory for the index per chromosome (42\% of its area).

\subsubsection{Execution Breakdown of \tgmasic{}}
\label{sec:breakdown}
\begin{figure}[h]
    \centering
    \includegraphics[width=0.9\columnwidth]{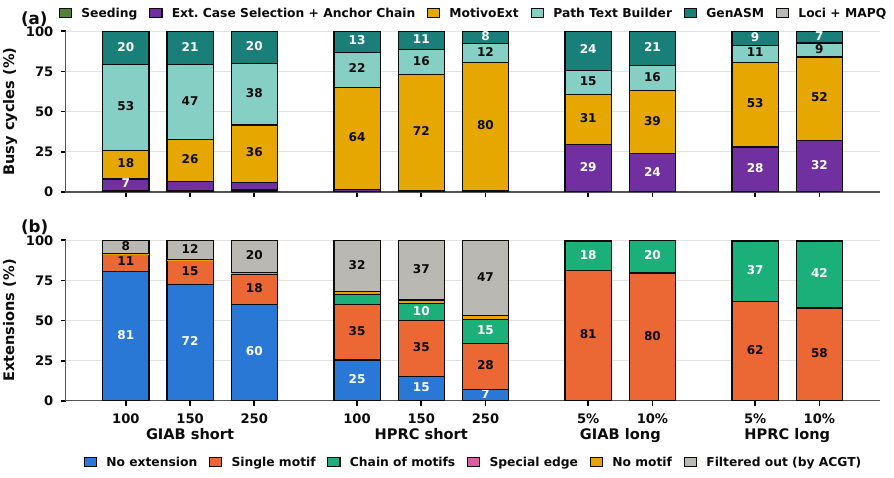}
    \caption{(a) Share of the busy cycles of a \tgm{} unit per module.  (b) How \tgmext{} resolves the extensions.}
    \label{fig:stage_breakdown}
\end{figure}
\fig{\ref{fig:stage_breakdown}} shows (a)~the share of the busy cycles of each module of a \tgmasic{} unit and (b)~how \tgmext{} resolves the extensions.

\noindent\textit{Short reads.} On the GIAB graphs, 60-81\% of the seed hits need no extension, i.e., their subgraph is a single node (\fig{\ref{fig:stage_breakdown}b}), so \tgmext{} takes less than 36\% of the busy cycles (\fig{\ref{fig:stage_breakdown}a}). \textcolor{black}{The largest module is the Path Text Builder (38-53\%), which fetches the node sequences of the candidate paths and builds the linear reference sequence to align.} For these seed hits, \tgmasic{} and SeGraM perform the same S2S alignment with the same algorithm, GenASM, which explains the modest gain of \tgmasic{} over SeGraM for short reads (\sect{\ref{sec:s2g-accel}}). On the denser HPRC graphs, \textcolor{black}{41-46\% of the seed hits need a single motif or a chain of motifs, and \tgmext{} filters out 32-47\% of the candidate subgraphs (\fig{\ref{fig:stage_breakdown}b}), so \tgmext{} dominates the busy cycles (64-80\%).} Due to this filtering, alignment is no longer the bottleneck and takes only 8-13\% of the busy cycles.

\noindent\textit{Long reads.} The chain of anchors splits each long read into short gaps that a single motif or a short chain of motifs resolves. \textcolor{black}{Thus, \tgmasic{} resolves each gap with a few motif lookups instead of a DP-based alignment over the gap's subgraph.} GenASM takes at most 24\% of the busy cycles, compared to nearly all of SeGraM's mapping time (\sect{\ref{sec:motivation}}). \tgmasic{} thus moves the bottleneck from DP-based alignment to lightweight motif-based extension, and its benefit grows with graph density.

\subsubsection{\textcolor{black}{Scalability}}
\label{sec:scalability}
\begin{figure}[h]
    \centering
    \includegraphics[width=0.9\columnwidth]{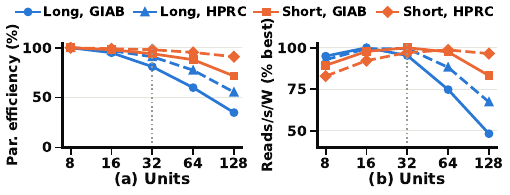}
    \caption{\textcolor{black}{Scalability of \tgm{} from 8 to 128 units on one HBM2E stack with the shared index. (a)~Parallel efficiency relative to 8 units and (b)~reads/s per Watt (compute units and HBM) relative to the best unit count of each dataset.}}
    \label{fig:scalability}
\end{figure}
\klgh{\fig{\ref{fig:scalability}a} and \fig{\ref{fig:scalability}b} show the throughput and energy efficiency, respectively, when scaling \tgm{} from 8 to 128 units on a single HBM2E stack. In this case, one index is shared across all channels. We assign one unit per channel for 8 units. We assign 1, 2, 4, and 8 units per pseudo-channel for 16, 32, 64, and 128 units, respectively. Energy efficiency measurements include the power of the units and of the HBM2E stack. 
We make two observations.
\textcolor{black}{First, the throughput scales almost linearly up to 32 units for short reads (93-99\% parallel efficiency) and for long reads on the HPRC graphs (90-92\%), and saturates beyond 32 units. For example, for long reads on the HPRC graphs, scaling from 32 to 64 units lowers the parallel efficiency from 90-92\% to 74-81\%. For long reads on the GIAB graphs, the parallel efficiency already drops to 79-83\% at 32 units.}
Second, 32 units provide the best balance between throughput and energy efficiency, e.g., across all datasets, 32 units reach 93-100\% of the best reads/s per Watt, whereas 128 units reach only 43-72\% for long reads.
Thus, we use 32 units to make the most of performance and energy efficiency from a single HBM2E stack. }

\subsection{Area, Power, and Energy}
\label{sec:energy}
\noindent\textbf{Area and Power.} \tab{\ref{tab:area-power}} shows the area breakdown of a \tgm{} unit at 28\,nm. A \klgh{single} unit occupies \textcolor{black}{1.146\,mm$^2$}, so the 32-unit accelerator occupies \textcolor{black}{36.7\,mm$^2$}\textcolor{black}{, compared with 27.7\,mm$^2$ for SeGraM and 7.93\,mm$^2$ for HARP. \tgmasic{} is thus larger than both accelerators and consumes more power than HARP (see below), but it provides higher throughput per mm$^2$ and per Watt than both (\sect{\ref{sec:s2g-accel}})}.
\revb{\labelb{3} We make two observations. First, \textcolor{black}{the largest share} of the unit's area implements \tgm{}'s motif-based mechanisms. \tgmext{} (\textcolor{black}{23.2\%}) and the extension case selection with the chain of anchors (\textcolor{black}{21.0\%}) together occupy \textcolor{black}{44.2\%} of the unit. The extension case selection module occupies a large area because, for long reads, it stores per-read seed hits and builds the chain of anchors.
Second, motif processing is inexpensive: \tgmext{} occupies \textcolor{black}{less than} a quarter of the unit.}
\begin{table}[h]
\centering
\caption{Area breakdown of a \tgm{} unit (28\,nm).}
\label{tab:area-power}
\footnotesize
\setlength{\tabcolsep}{3pt}
\renewcommand{\arraystretch}{0.9}
\begin{tabular}{lr}
\hline
\textbf{Module} & \textbf{mm\textsuperscript{2} (\%)} \\
\hline
Minimizer Extraction + Seeding & \textcolor{black}{0.053 \phantom{0}(4.5)} \\
Extension Case Selection + Anchor Chain & \textcolor{black}{0.241 (21.0)} \\
\tgmext{} & \textcolor{black}{0.265 (23.2)} \\
S2S Alignment and Loci & \textcolor{black}{0.490 (42.8)} \\
\quad incl.\ GenASM (DC + TB)~\cite{cali2020genasm} & 0.334 \textcolor{black}{(29.2)} \\
Shared Read Buffer + Pipeline Control & \textcolor{black}{0.097 \phantom{0}(8.5)} \\
\hline
\textbf{Total} & \textbf{\textcolor{black}{1.146} (100)} \\
\hline
\end{tabular}
\end{table}

The power of a \tgm{} unit is \textcolor{black}{0.33\,W} for short reads and \textcolor{black}{0.376\,W} for long reads. The 32-unit accelerator consumes \textcolor{black}{10.6\,W} and \textcolor{black}{12.0\,W}, respectively, compared with 24.3\,W for SeGraM\textcolor{black}{ and 3.82\,W for HARP}. \textcolor{black}{The HBM2E memory adds 1.2-7.5\,W, depending on the read length and graph type.} 
\begin{figure}[h]
    \centering
    \includegraphics[width=0.9\columnwidth]{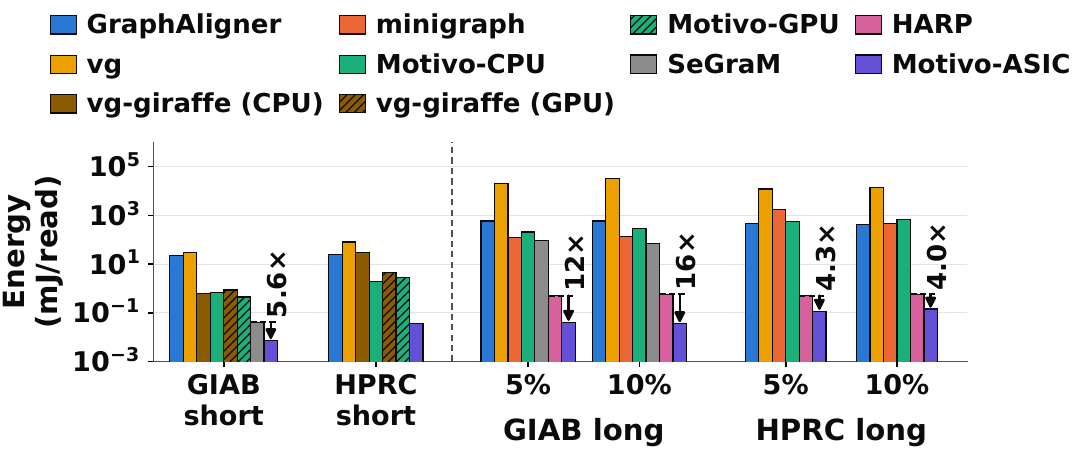}
    \caption{Energy per read on the GIAB and HPRC graphs. Labels show \tgmasic{}'s energy reduction over SeGraM (short reads) and HARP (long reads). For short reads, the bars average the energy per read over the three read lengths (5.6$\times$ vs.\ SeGraM); averaging reads/s per Watt instead, as in \fig{\ref{fig:iso_area_power}}, gives 5.3$\times$.}
    \label{fig:energy}
\end{figure}

\noindent\textbf{Energy.} \fig{\ref{fig:energy}} shows the energy per read of all systems. \tgm{} consumes the least energy per read for all datasets and types of graphs\textcolor{black}{. Compared with the S2G accelerators, \tgm{} consumes 5.3$\times$ less energy per read than SeGraM for short reads on the GIAB graphs, and 4.0-4.3$\times$ less than HARP for long reads on the HPRC graphs}. \textcolor{black}{\fig{\ref{fig:energy}} shows the compute energy of each system. If we also include the energy of \tgm{}'s HBM2E memory, the entire \tgm{} system still consumes less energy per read than the compute part alone of every other system, i.e., 4.8$\times$ less than SeGraM for short reads on the GIAB graphs, and 3.1-3.2$\times$ less than HARP for long reads on the HPRC graphs.}

\subsection{Accuracy}
\label{sec:accuracy}
We use simulated reads to determine the reads' ground-truth mapping and use it to measure \tgm{}'s accuracy. We consider a read \emph{correctly mapped} if \inum{i} its reported start and end positions on the graph are the same as its true origin, and \inum{ii} its reported alignment (CIGAR) has at most $E$ edits (\sect{\ref{sec:back-e}}). Overall, \textcolor{black}{\tgm{} achieves 99.1\% (99.7\%) accuracy for short (long) reads on the GIAB graphs and 98.0\% (96.8\%) on the denser HPRC graphs. The state-of-the-art S2G mappers GraphAligner and vg report 96.6\% and 93.8\% accuracy, respectively~\cite{rautiainen2020graphaligner,siren2021pangenomics}.}

\begin{figure}[h]
\centering
\includegraphics[width=0.9\columnwidth]{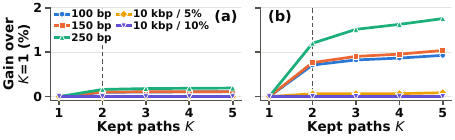}  
\caption{Accuracy gain when aligning $K$ paths instead of a single one for (a) GIAB and (b) HPRC graphs.}
\label{fig:topk}
\end{figure}
\revcqone{\textbf{Top-$K$ Paths Analysis.} 
\fig{\ref{fig:topk}} shows the accuracy gained when scaling the number $K$ of extended paths. On the GIAB graphs, a single path ($K=1$) already achieves around 99\% accuracy, and a second path adds only 0.1-0.2\%. On the denser HPRC graphs, a second path improves short-read accuracy by 0.7-1.2\%, increasing with read length, while long reads gain at most 0.1\%. We therefore keep $K=2$ paths, since aligning $K>2$ paths increases alignment operations by up to 2.5$\times$ (4.3$\times$) for short (long) reads.}

\revcqthree{\revmid{cqthree}{CQ3}\textbf{Contribution of Each Mechanism.} We evaluate the contribution of each mechanism in two complementary ways (\fig{\ref{fig:ablation_acc}}). 
\begin{figure}[h]
\centering
\includegraphics[width=0.9\columnwidth]{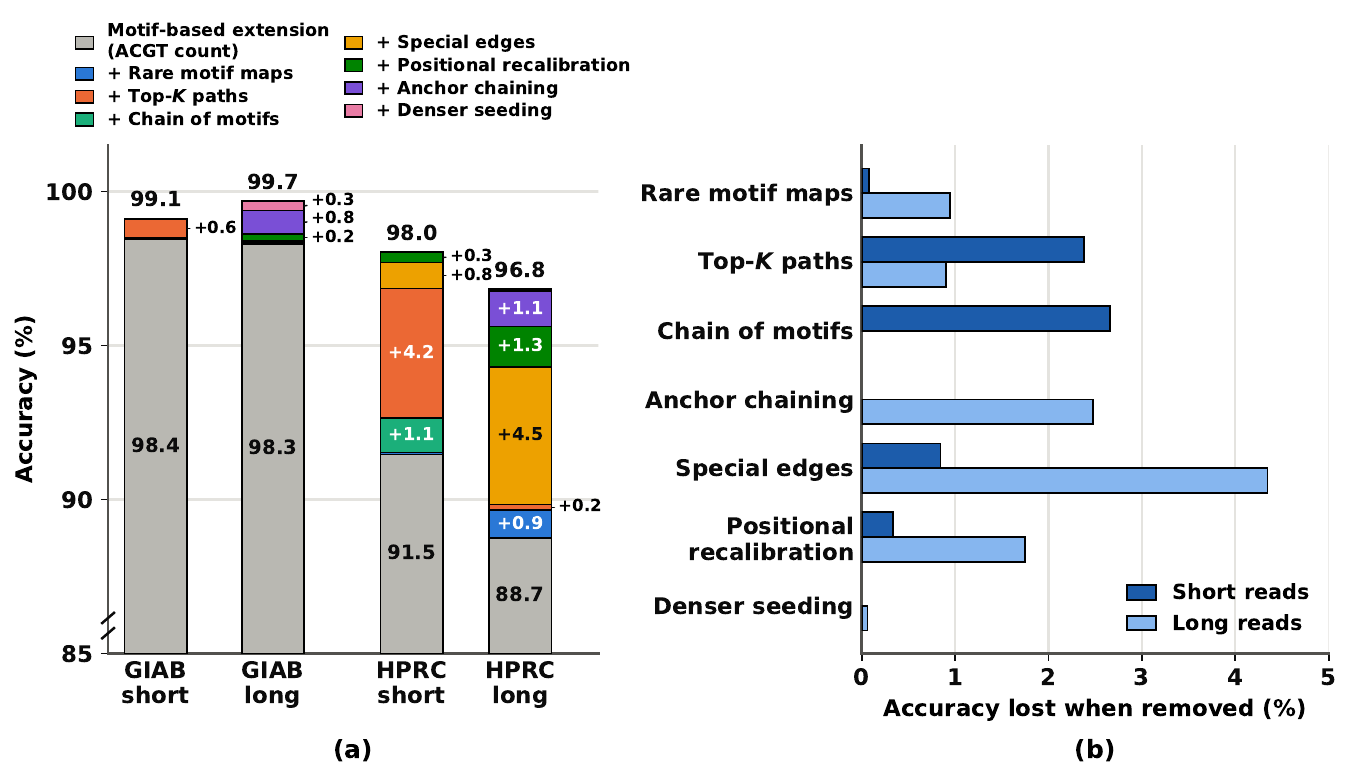}   \caption{(a) Accuracy contribution of each mechanism.  (b) Accuracy loss from individual mechanism removal (HPRC).}
\label{fig:ablation_acc}
\end{figure}
\revmid{rf}{F2}First, \fig{\ref{fig:ablation_acc}a} shows how accuracy builds up as we cumulatively add the mechanisms to our original design (i.e., the one that keeps a single path, stores only the 256 on-chip motif groups, and uses neither special edges nor positional recalibration). On the GIAB graphs, the original design already achieves almost perfect accuracy (98.3-98.4\%), confirming that motif topology and base composition alone are highly accurate for less complex graphs. On the HPRC graphs, the mechanisms add 6.6\% (8.1\%) for short (long) reads.    
Second, to evaluate the effect of each mechanism on the accuracy in isolation, \fig{\ref{fig:ablation_acc}b} removes one mechanism at a time from the final design and shows the accuracy loss for the denser HPRC graphs. Without positional recalibration or special edges, the design loses up to 4.3\% accuracy on denser graphs. On the GIAB graphs, removing any single \tgmext{} mechanism costs at most 0.3\% (0.2\%) for short (long) reads. Thus, for less complex graphs, \tgm{} maps reads with high accuracy even with $K=1$ and without the additional mechanisms.}

\section{Related Work}
\label{sec:related-work}

\konkanello{\tgm{} is a novel algorithm-architecture co-designed sequence-to-graph mapping system that exploits the topological recurrence observed in genome graphs.}

\head{Accelerating S2S Mapping}
Many works accelerate sequence-to-sequence (S2S) mapping, i.e., mapping reads to a single linear reference genome~\cite{alser2020accelerating,alser2020technology,alser2022molecules,mutlu2023accelerating,sadasivan2024genomic}, with algorithmic and hardware optimizations for each step of the mapping pipeline.
\textbf{Seeding} is accelerated with hash-, minimizer-, syncmer-, strobemer-, and FM-index-based seeds and with optimized seed selection~\cite{Langmead_2009,langmead2012fast,li2013aligningsequencereadsclone,li2018minimap2,hach_mrsfast_2010,weese_razersfast_2009,marccais2017improving,jain_weighted_2020,zheng_improved_2020,Edgar_2021,Sahlin_2021,sahlin2022strobealign,xin2016optimal,xin_context-aware_2020,Firtina_2023a,ekim_efficient_2023,vasimuddin2019efficient,Jung_2022} (also for raw nanopore signals~\cite{firtina2023rawhash,firtina2023rawhash2,lindegger2023rawalign}), and with hardware for FM-index search and seed lookup on FPGAs, ASICs, and processing-in/near-memory (PIM/PNM) systems~\cite{waidyasooriya2015hardware,subramaniyan2021accelerated,fujiki2018genax,li2021pim,zokaee2018aligner,angizi2019aligns,khatamifard2021genvom,laguna2020seed,chen2023gem}.
\textbf{Pre-alignment filtering} quickly rejects dissimilar read-reference pairs before alignment, in software~\cite{xin2013accelerating,xin2015shifted,alser2017magnet,alser2020sneakysnake} and in hardware on FPGAs~\cite{alser2017gatekeeper,alser2019shouji,alser2020sneakysnake,singh2021fpga}, GPUs~\cite{bingol2021gatekeeper,alser2020sneakysnake}, and PIM/PNM and in-storage systems~\cite{kim2018grim,nag2019gencache,kaplan2018rassa,hameed2021alpha,He2024shiftcam,mao2022genpip,mansouri2022genstore}.
\textbf{Chaining} of seed hits relies on sparse DP algorithms~\cite{myers_chaining_1995,abouelhoda_chaining_2005,otto_fast_2011,makinen_chaining_2020,jain_co-linear_2022,shaw_proving_2023,li2018minimap2} and is accelerated on CPUs with SIMD~\cite{kalikar_accelerating_2022}, on GPUs~\cite{guo2019hardware,sadasivan2023accelerating,dong2024mm2}, on FPGAs~\cite{guo2019hardware,liyanage2023efficient}, with custom RISC-V instructions~\cite{liyanage_accelerating_2024}, and with programmable DP accelerators~\cite{gu2023gendp}.
\textbf{Alignment} is accelerated algorithmically with banded, bit-parallel, difference-recurrence, wavefront, and A*-based formulations of the classic DP algorithms~\cite{needleman1970general,smith1981identification,gotoh1982improved,ukkonen_algorithms_1985,myers1999fast,vsovsic2017edlib,Zhao_2013,suzuki_introducing_2018,marco2021fast,marcosola2023optimal,groot_koerkamp_exact_2024,grootkoerkamp2024apa2,cali2020genasm,lindegger_algorithmic_2022,Lindegger2023scrooge,rumpf_sequencelab_2023}, and in hardware on GPUs~\cite{liu2009cudasw++,liu2010cudasw++,liu2013cudasw++,liu2015gswabe,de2016cudalign,nishimura2017accelerating,ahmed2019gasal2,zeni2020logan,ahmed_gpu_2020,goenka_segalign_2020,aguado-puig_wfa-gpu_2022,park_saloba_2022,Park2024agatha,zeni2024leveraging}, on FPGAs~\cite{fei2018fpgasw,rucci2018swifold,li2021pipebsw,liao_adaptively_2018,haghi2021fpga,haghi_wfa-fpga_2023,teng_adapting_2023,fujiki2020seedex}, with ASICs and ISA extensions~\cite{madhavan2014race,turakhia2018darwin,Turakhia2019darwinwga,fujiki2018genax,cali2020genasm,gu2023gendp,Walia2024talco,pavon_quetzal_2024,doblas2025smx}, and with PIM~\cite{huangfu2018radar,gupta2019rapid,kaplan2020bioseal,angizi2020pim,Zhang_2023_alignerD,Diab2023,AlonsoMarn2024bimsa,Mognol2024}.
Other works accelerate complete S2S mappers~\cite{goyal2017ultra,houtgast2017efficient,houtgast2018hardware,chen2014accelerating,chen2021high,banerjee2018asap,wilton2015arioc,cheng2018bitmapper2,ham2020genesis,ham2021accelerating,liyanage2023efficient,Cheng2024mm2hwaccel,Sun2025,eudine2026genpairx} and the remapping of reads between reference genomes~\cite{kim2024airlifttcbb,kim_fastremap_2022}.
All these works target a linear reference and do not handle the complexities of genome graphs.\\
\head{Software Tools for S2G Mapping}
S2G mapping tools~\cite{garrison2018variation,rautiainen2020graphaligner,Li2020minigraph,siren2021pangenomics,dvorkina2020spaligner,liu2016debga,monti2026faster} improve \textbf{seeding}~\cite{schneeberger2009simultaneous,heydari2018browniealigner,ivanov2022fast,ma2023chaining,sahlin2022strobealign,li2012exploring}, \textbf{indexing}~\cite{burrows1994block,huang2013short,siren2014indexing,ferragina2000opportunistic,rossi2022moni,siren2020haplotype}, \textbf{filtering}~\cite{limasset2016read,li2018minimap2,chandra2023sequence}, and \textbf{extension}~\cite{marco2021fast,zhang2022fast,Ivanov2020,ivanov2022fast}, but still perform redundant, expensive DP-based alignment (\sect{\ref{sec:motivation}}). \tgm{} avoids it with a motif-based index and a topology-aware extension mechanism.\\
\head{Hardware Acceleration of Graph-Based Genome Analysis}
Prior works accelerate S2G mapping with SIMD~\cite{jain2019pasgal,Darby2020vargas}, GPUs~\cite{Feng2021,zhu2025parabricks}, ASICs~\cite{Cali2022SeGraM,Zhang2024}, processing-in-memory (PIM)~\cite{Huang2023meg2}, and in-storage processing~\cite{mansouri2026grains-extended}. However, SeGraM~\cite{Cali2022SeGraM}, the state-of-the-art hardware accelerator for S2G mapping that maps both short and long reads, lacks filtering of redundant computations and support for emerging dense pangenome graphs. HARP~\cite{Zhang2024} exploits the long sequential regions of the graph to accelerate indexing and alignment, but supports only long reads and performs seeding on the host. MeG2~\cite{Huang2023meg2} maps SeGraM's seeding and bitwise alignment to PIM. GRAINS~\cite{mansouri2026grains-extended} targets genome graphs that exceed main memory and reduces their storage I/O with in-storage and in-flash processing, forwarding the candidate graph regions it identifies to SeGraM.
MeG2 and GRAINS thus accelerate the work by moving it closer to the data~\cite{mutlu2019processing,singh2018review,singh_napel_2019,singh_nero_2020,singh2021fpga}. 
\revb{\labelb{5} None of these works exploits the recurring local topologies of genome graphs. \tgm{} does, reducing S2G mapping to a few S2S alignments per read; it accurately maps both short and long reads and runs the entire pipeline on chip, from seeding to mapping quality.} This algorithmic reduction is orthogonal to where the computation runs, as shown in \sect{\ref{sec:evaluation-analyses}}. The \tgm{} algorithm is not tied to one platform, and \tgmext{} needs only lookups, bitmask operations, and additions, which suit in-memory and in-storage logic. \tgm{} can therefore be deployed in a MeG2-style PIM system, and can replace SeGraM as the alignment engine of GRAINS, combining GRAINS's I/O savings with far fewer alignments per read.\\
\head{Graph Accelerators for Other Applications}
\reva{\labela{3}Graph-based approximate nearest neighbor search (ANNS) accelerators~\cite{ndsearch,proxima} also index offline and filter out unpromising candidates, but they return the approximate nearest stored fixed-length vectors: they cannot shift for indels and do not store candidate paths. Using them for S2G mapping requires flattening the graph into fixed-length windows, which either ignores allele combinations or enumerates them into a very large index, and still requires alignment.}

\section{Conclusion}
\label{sec:conclusion}
We propose \tgm{}, the first algorithm-architecture co-design for S2G mapping that exploits recurring subgraph topologies, based on the key observation that a genome graph consists of small subgraphs that share a limited set of motifs. \tgm{}'s motif-based algorithm identifies, before alignment, the few paths that can lead to a valid alignment, replacing graph traversals and DP with memory lookups and arithmetic operations. It enables high-performance execution on CPUs and GPUs, and high-performance, low-energy execution on the co-designed \tgmasic{}.

\balance
\bibliographystyle{unsrt-keepcase}
\bibliography{ref,akmami,refs_grains}
\end{document}